\documentclass[aps,prc,twocolumn,showpacs,groupedaddress]{revtex4}
\usepackage{graphicx} 
\usepackage{subfigure}
\usepackage{amsfonts, amssymb, latexsym}
\usepackage{epsfig}

\begin{document}



\title{Proton elastic form factor ratios to $Q^2$ = 3.5 GeV$^2$ by
polarization transfer}

\author {
V.~Punjabi,$^{1,*}$
C.F.~Perdrisat,$^{2}$
K.A.~Aniol,$^{7}$
F.T.~Baker,$^{4}$
J.~ Berthot,$^{6}$
P.Y.~ Bertin,$^{6}$ 
W.~ Bertozzi,$^{22}$
A.~ Besson,$^{6}$
L.~Bimbot,$^{26}$
W.U.~Boeglin,$^{10}$
% ...
}

\vspace{0.1in}

\affiliation{
\baselineskip 2 pt
\vskip 0.3 cm
{\rm The Jefferson Lab Hall A Collaboration} \break
\vskip 0.1 cm
$^{1}$Norfolk State University, Norfolk, VA 23504\\
$^{2}$College of William and Mary, Williamsburg, VA 23187\\
% ...
$^{4}$University of Georgia, Athens, GA 30602\\
% ...
$^{6}$Universit\'{e} Blaise Pascal/CNRS-IN2P3, F-63177 Aubi\`{e}re, France\\ 
$^{7}$California State University, Los Angeles, Los Angeles, CA 90032\\
% ...
$^{10}$Florida International University, Miami, FL 33199\\
% ...
$^{22}$Massachusetts Institute of Technology, Cambridge, MA 02139\\
% ...
$^{26}$Institut de Physique Nucl\'{e}aire, F-91406 Orsay, France\\
% ...
}

\date{\today}

\altaffiliation{ {\it {Email address:punjabi@jlab.org}} \\
$^a$ Present address: Christopher Newport University, Newport News, VA 23606, USA.\\
$^b$ Present address: University of Lund, Box 118, SE-221 00 Lund, Sweden.\\
$^c$ Present address: California Institute of Technology, Pasadena, CA 91125, USA.\\
$^d$ Present address: North Carolina Ag. and Tech. State University, Greensboro, NC 27411, USA.\\ 
$^e$ Present address: Renaissance Technology Corp., Setauket, NY 11733, USA.\\
$^f$ Present address: Saint Mary's University, Halifax, NS, Canada B3H 3C3.\\
$^g$ Present address: George Washington University, Washington, DC 20052, USA.\\
$^h$ Present address: NIKHEF, Amsterdam, The Netherlands.\\
$^i$ Present address: Carnegie Mellon University, Pittsburgh, PA 15217, USA.}

\begin{abstract}
The ratio of the proton elastic electromagnetic form factors, $G_{Ep}/G_{Mp}$, was 
obtained by measuring $P_{t}$ and $P_{\ell}$, the transverse and longitudinal 
recoil proton polarization components, respectively, 
for the elastic $\vec e p \rightarrow e\vec p$ ~reaction in the four-momentum transfer squared range  
of 0.5 to 3.5~GeV$^2$. In the single-photon exchange approximation, the ratio $G_{Ep}/G_{Mp}$ 
is directly proportional to the ratio 
$P_t/P_{\ell}$. The simultaneous measurement of $P_{t}$ and $P_{\ell}$ in a polarimeter 
reduces systematic uncertainties. The results for the ratio $G_{Ep}/G_{Mp}$ 
show a systematic decrease with increasing $Q^2$, indicating  for the 
first time a definite difference in the distribution of charge and magnetization 
in the proton. The data have been re-analyzed and systematic uncertainties have become 
significantly smaller than previously published results.  
\end{abstract}

\maketitle

\section{INTRODUCTION}

One of the fundamental goals of nuclear physics is to understand the 
structure and behavior of strongly interacting matter in terms of its basic
constituents, quarks and gluons. An important step towards this goal is  
the characterization of the internal structure of the nucleon; the 
four Sachs elastic electric and magnetic form factors of the proton and neutron, 
$G_{Ep}$, $G_{Mp}$, $G_{En}$ and $G_{Mn}$, are 
key ingredients of this characterization. The elastic electromagnetic form factors 
are directly related to the charge and current distributions inside the nucleon; 
these form factors are among the most basic observables of the nucleon. 
 
The first direct evidence that the proton has an internal structure came from a measurement of  
its anomalous magnetic moment 70 years ago by O. Stern \cite{stern}; it is 2.79 times 
larger than that of a Dirac particle of the same mass. The first measurement of the charge 
radius of the proton, by Hofstadter {\it{et al.}} \cite{hofstader}, yielded a value of 0.8 fm,  
quite close to the modern value. 

The theory that describes the strong interaction between quarks and gluons 
is Quantum Chromodynamics (QCD). Perturbative QCD (pQCD) 
makes rigorous predictions when the four-momentum transfer squared, $Q^2$, is very large and 
the quarks become asymptotically free.  
It is not known precisely at what value of $Q^2$ pQCD may start to dominate; however, expectations 
are that this will not occur until $Q^2$ is at least in the tens of GeV$^2$ \cite{Llew}. 
Predicting nucleon form factors in the non-perturbative regime, where soft scattering 
processes are dominant, is very difficult. As a consequence there 
are many phenomenological models which attempt to explain the data in this domain;  
precise measurements of the nucleon form factors are necessary to constrain 
and test these models. Only the magnetic form factor 
of the proton, $G_{Mp}$, is known with very good accuracy in this region. 
The electric form factor, $G_{Ep}$, was not well measured beyond $Q^2$ of 
1~GeV$^2$ before this experiment. Both $G_{En}$ and $G_{Mn}$, the electric and 
magnetic form factors 
of the neutron, respectively, were also poorly known at any $Q^2$ value until recently.
New measurements of $G_{Mn}$ at Jefferson Lab \cite{brooks} up to $Q^2$=4.8~GeV$^2$
will bring the knowledge of this form factor to 
comparable levels of accuracy as for $G_{Mp}$. For the 
neutron electric form factor, two new Jefferson Lab experiments \cite{hallC,gen} 
have extended the $Q^2$ range to 1.5~GeV$^2$, and two approved experiments \cite{bogdan,madey}
will soon extend the $Q^2$ 
range to 4.3~GeV$^2$, with an accuracy comparable to that of the three other form factors.

The electromagnetic interaction provides a unique tool to investigate 
the internal structure of the nucleon. The measurement of electromagnetic 
form factors in elastic, inelastic, and structure functions in deep inelastic scattering of electrons and 
muons, has been a rich source of information on the structure of the nucleon. 

In the single virtual photon exchange approximation for elastic scattering, the hadron 
current operator can be expressed in terms of two form factors: $F_{1}$, the Dirac
form factor, and $F_{2}$, the Pauli form factor. These form factors and the Sachs electric 
and magnetic form factors are related according to:
\begin{equation}
G_{E}=F_{1}-\tau \kappa F_{2} \mbox{ and } G_{M}=F_{1}+\kappa F_{2},
\label{eq:sach_f1f2}
\end{equation}
\noindent
where $\tau = Q^2/4M_{p}^{2}$, $\kappa$ is the anomalous magnetic moment and $M$ the mass of the proton. 
In the limit $Q^2\rightarrow~0$, $G_{Ep}=1$, $G_{En}=0$, $G_{Mp}=\mu_p$, and $G_{Mn}=\mu_n$, 
where $\mu_p$ and $\mu_n$ are the nucleon magnetic moments. In the Breit frame, $G_{E}$ and $G_{M}$ are the Fourier 
transforms of the charge and magnetization distributions in the nucleon, respectively.

\subsection{Previous $G_{Ep}$ Measurements Using the Rosenbluth Separation Method}

Both the elastic cross section and the polarization observables of the elastic $ep$ reaction
can be expressed in terms of either the Sachs or the Dirac and Pauli form 
factors. These form factors are Lorentz scalars and depend only upon $Q^2$, 
the four-momentum transfer squared of the reaction. A complete 
separation of the electric and magnetic terms is evident in the cross section formula
when the Sachs form factors are used. It is then possible to
obtain both $G_{Ep}^2$ and $G_{Mp}^2$ separately, using the Rosenbluth 
method \cite{rosenbluth,hand}. In the one-photon exchange approximation, the cross section in terms of the
Sachs form factors can be expressed as:
\begin{equation}
\frac{d\sigma }{d\Omega } =  \frac{{\alpha}^2~E_e~\cos^2\frac
{\theta_e}{2}}{4E_{beam}^3~\sin^4\frac{\theta_e}{2}}  
\left[ G_{Ep}^2+\frac{\tau}{\epsilon}G_{Mp}^2 \right]\left (\frac{1}{1+\tau }\right ),
\label{eq:xngegm}
\end{equation}
\noindent
where $\epsilon=\left[1+2(1+\tau)\tan^2(\frac{\theta_e}{2})\right]^{-1}$
is the longitudinal polarization of the virtual photon, with values between 0$<\epsilon<$1, 
$E_{beam}$ and
$E_{e}$ are the energies of the incident
and scattered electron, respectively, and
$\theta_e$ is the electron scattering angle in the laboratory frame. 

Figure 1 shows previous results of $G_{Ep}$ and  
$G_{Mp}$ obtained by Rosenbluth separations, plotted as the ratios 
$G_{Ep}/G_D$ and $ G_{Mp}/\mu_{p} G_D$ versus $Q^2$,
up to 6~GeV$^2$. Here $G_D=(1 + Q^2/m_D^2)^{-2}$ is 
the dipole form factor, with the constant $m_D^2$ empirically determined 
to be 0.71~GeV$^2$. For $Q^2 <$ 1~GeV$^2$, the uncertainties for both 
$G_{Ep}$ and $G_{Mp}$ are only a few percent, and
one finds that $G_{Mp}/ \mu_{p} G_D \simeq G_{Ep}/G_D \simeq 1 $. 
For $G_{Ep}$ above $Q^2$ = 1~GeV$^2$, the large uncertainties 
and the scatter in results between different 
experiments, as seen in Fig.~1, illustrate the difficulties in obtaining 
$G_{Ep}$ by the Rosenbluth separation method. In contrast, the uncertainties 
for $G_{Mp}$ obtained from cross section data with the assumption $G_{Ep}=G_{Mp}/\mu_p$, 
remain small up to $Q^2$ = 31.2~GeV$^2 $\cite{sill}.
In Eq.~(\ref{eq:xngegm}) the $G_{Mp}$ part of the cross section, which is about 
$\mu_p^2$ times larger than the $G_{Ep}$ part, is also multiplied by $\tau$; therefore, as $Q^2$ 
increases, the cross section becomes dominated by the $G_{Mp}$ term, making the extraction of $G_{Ep}$ 
more difficult by the Rosenbluth separation method.  

\begin{figure}
\begin{center}
\epsfig{file=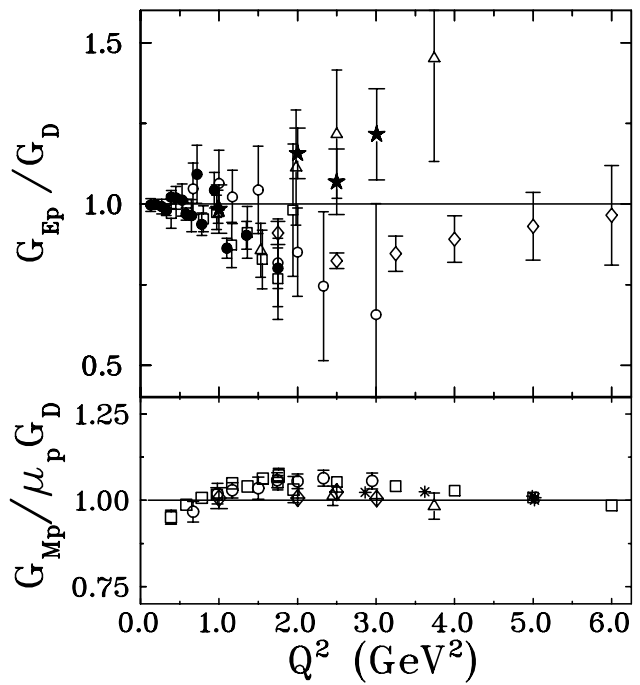,width=3.25in}
\vspace{0.1in}
\caption [] { World data prior to 1998 for (a) $G_{Ep}/G_D$ and 
(b) $G_{Mp}/\mu_p G_D$ versus $Q^2$.
Refs. Litt {\it et al.} \cite{litt}$\triangle$, Berger {\it et al.} \cite{berger}$\Box$,  
Price {\it et al.} \cite{price}$\bullet$, Bartel {\it et al.} \cite{bartel}$\circ$, Walker 
{\it et al.} \cite{walker}$\star$, Andivahis {\it et al.} \cite{andivahis}$\Diamond$ and Sill 
{\it et al.} \cite{sill}$\ast$.}
\label{fig:gepgd_gmpgd}
\end{center} 
\end{figure}

\subsection {Polarization Transfer Method }

The proton form factor ratio $G_{Ep}/G_{Mp}$ can be obtained 
from polarization observables of the $\vec ep \rightarrow e\vec p$ or $\vec e \vec p \rightarrow ep$ 
reaction, the recoil proton polarization transfer 
coefficients or the beam-target polarization asymmetry, respectively. Both reactions contain an 
interference term proportional to $G_{Ep}G_{Mp}$; hence
polarization experiments are able to obtain the electric form factor $G_{Ep}$ even when it is very small.

For one-photon exchange, in the $\vec ep\rightarrow e\vec p$ reaction, the 
scattering of longitudinally 
polarized electrons results in a transfer of polarization to the recoil proton 
with only two non-zero components, $P_{t}$ perpendicular to, and $P_{\ell }$  
parallel to the proton momentum in the scattering plane.  For 100 \% longitudinally polarized 
electrons, the polarizations are \cite{akh1,dombey,akh2,arnold}: 

\begin{eqnarray}
I_{0}P_{n}&=&0
\label{eq:pn} \\
I_{0}P_{t}&=&-2\sqrt{\tau (1+\tau )}G_{Ep}G_{Mp}\tan \frac{\theta_{e}}{2}
\label{eq:pt} \\
I_{0}P_{\ell}&=&\frac{1}{M_p}(E_{beam}+E_{e})\sqrt{\tau (1+\tau )}
G_{Mp}^{2}\tan ^{2}\frac{\theta _{e}}{2}  
\label{eq:pl}
\end{eqnarray}
\noindent
where $I_0$ is proportional to the unpolarized cross section and is given by:

\begin{equation}
I_{0}=G_{Ep}^{2}+\frac{\tau}{\epsilon} G_{Mp}^{2}
\label{eq:io}
\end{equation}

Eqs. (\ref{eq:pt}) and ~(\ref{eq:pl}) show that $I_oP_t$ and $I_oP_{\ell}$ are 
proportional to $G_{Ep}G_{Mp}$ and $G_{Mp}^2$, respectively. Together 
these equations give:

\begin{equation}
\frac{G_{Ep}}{G_{Mp}}=-\frac{P_{t}}{P_{\ell}}\frac{(E_{beam}+E_{e})} 
{2M_p}\tan \frac{\theta_{e}}{2}
\label{eq:ratio}
\end{equation}

If only the polarization components $P_t$ and $P_{\ell}$ are measured, 
as was the case in this experiment, 
then from Eqs. (\ref{eq:pt}) and ~(\ref{eq:pl}) the form factors $G_{Ep}$ and $G_{Mp}$ cannot 
be obtained separately, only their ratio can be determined. To obtain $G_{Ep}$ and $G_{Mp}$ separately,
$I_{0}$ in Eq. (\ref{eq:io}) must be obtained from cross section measurements.  

The ratio $G_{Ep}/G_{Mp}$ is obtained from a single measurement of 
the two recoil polarization components $P_t$ and $P_{\ell}$ in a polarimeter, whereas  
the Rosenbluth method 
requires at least two cross section measurements made at different energy 
and angle combinations at the same $Q^2$. 

The recoil polarization method was first used in electron scattering experiments to 
obtain the neutron form
factors in the $^{2}H(\vec{e},e' \vec{n})p$
reaction \cite{eden} and to measure the form factor ratio $G_{Ep}/G_{Mp}$ 
in $\vec{e}p\rightarrow e\vec{p}$ ~for the free proton \cite{milbrath,pospischil}, 
as well as in  the $^{2}H(\vec{e},e' \vec{p})n$ 
reaction for the proton in 
the deuteron at small $Q^2$-values \cite{barkhuff}. 

For completeness we mention here that a small normal component $P^{ind}_{n}$ is induced 
by two-photon exchange mechanism, independent of beam polarization. The observables 
$P_t$ and $P_{\ell}$ of this experiment are entirely due
to polarization transfer, and the analysis method described in this paper allows complete 
separation of helicity dependent and helicity independent polarization components. 

In this paper, we present the $G_{Ep}/G_{Mp}$ ratios, primary results from this experiment, 
obtained at Jefferson Lab using the recoil polarization method described here.
The experimental setup, in particular the focal plane polarimeter (FPP), 
is described in part II. The data analysis 
is presented in part III; this part also includes a discussion of the FPP calibration, the secondary 
results of the experiment, which are independent measurements of analyzing powers at ten proton 
energies between 0.244~GeV and 1.795~GeV. Part IV includes the main results of the 
experiment: the ratios $G_{Ep}/G_{Mp}$ at 0.5~GeV$^2 \leq Q^2\leq$ 3.5~GeV$^2$, and an analysis 
of systematic uncertainties.
A discussion of theoretical calculations as compared to the data is in part V and conclusions 
are presented in part VI. 

\section{THE EXPERIMENT}

The combination of high energy, current, polarization, and duty factor, unique to the 
Continuous Electron Beam Accelerator Facility (CEBAF) of the 
Thomas Jefferson National Accelerator Facility (JLab), makes it possible to investigate the
internal structure of the nucleon with higher precision than ever before. 
In this experiment, we have measured the 
polarization transferred to the recoil proton, with a longitudinally 
polarized electron beam scattered by an unpolarized hydrogen target. 

The experiment was performed in Hall A at JLab. The longitudinal 
and transverse polarizations of the outgoing proton were measured for the 
$\vec ep\rightarrow e\vec p$ reaction, in a range of $Q^2$ from 0.5~GeV$^2$ to 3.5~GeV$^2$. 
The beam energy ranged from 0.934~GeV to 4.091~GeV. For the five highest 
$Q^2$ data points, a bulk GaAs photo-cathode excited by circularly polarized 
laser light produced beams with polarization of  $\sim$0.39 and currents up to 
$\sim$115 $\mu$A; the sign of the beam helicity was changed at the rate of 30 Hz. 
For the lower $Q^2$ data points, a strained GaAs crystal 
was used and typical polarizations of $\sim$0.6 were achieved with 
currents between 5 $\mu$A and 15 $\mu$A; the sign of the beam helicity was changed at the rate of 1 Hz. 
The beam polarization was measured periodically with a Mott polarimeter in the injection line, 
and a M{\o}ller polarimeter \cite{glamazdin} in Hall A \cite{nimhallA}.

The Hall A M{\o}ller polarimeter uses magnetized 
ferromagnetic supermendure foils as a polarized electron target. The scattered 
electrons are detected in coincidence in the M{\o}ller spectrometer in the range of 
$75^{\circ}<\theta_{CM}<105^{\circ}$. The M{\o}ller spectrometer 
consists of three quadrupoles and a dipole magnet to bend scattered 
electrons toward the detector. The detector contains two identical modules for 
coincidence measurements; each module consists of a plastic scintillator and 
four blocks of lead glass.  The  M{\o}ller scattering cross section depends on 
the beam and the M{\o}ller target polarizations, $P_{e,i}$ and $P_{targ,i}$, respectively, 
$\sigma~\propto~(1~+~\sum_{i=X,Y,Z}(A_{ii}~P_{targ,i}~ P_{e,i}))$, 
where $i=X,Y,Z$ defines the projections 
of polarization. The analyzing power $A_{ii}$ depends on the scattering angle 
$\theta_{CM}$ and has its maximum at $\theta_{CM} = 90^{\circ}$. 
Statistical uncertainty varies between 0.2 \% and 0.8 \% for each measurement.
Total relative uncertainty of the beam polarization measurement is 
$\leq$~3 \%, when systematic and statistical uncertainties are combined.
In this experiment, the beam helicity cancels in the ratio $P_t/P_{\ell}$, so strictly 
speaking measurement of the beam polarization is not needed, but the beam polarization 
was measured periodically to ensure that the beam was polarized and also to allow FPP 
calibration as explained in section III D.

The beam current was monitored continuously during the experiment using 
resonant (RF) cavities. The beam current monitor (BCM) in Hall A consists 
of an Unser \cite{unser} monitor sandwiched between two RF cavities. The Unser monitor 
provides an absolute measurement of the current; the RF cavities are 
calibrated relative to the Unser monitor periodically. Both components 
are enclosed in a box to shield them from stray magnetic fields 
and for temperature stabilization. 

Beam position and direction at the target were determined from two beam 
position monitors (BPM) located at a distance of 7.524 m and 1.286 m upstream 
of the target position during this experiment. Each BPM is a cavity with a four-wire antenna with wires 
positioned at $\pm 45^{\circ}$ from the horizontal and vertical. 
The relative position of the beam on the target 
can be determined to about 100 $\mu $m for currents above 1 $\mu$A by 
using the technique of difference-over-sum between the signals from 
the four antenna wires. To obtain the absolute 
position of the beam, the BPMs are calibrated with respect to wire 
scanners which are located close to each of the BPMs at 7.353 m and 
1.122 m from the target. The wire scanners are surveyed with respect to the hall 
coordinates. The beam position is recorded for every event.

The cryogenic target contained three loops. Each loop included  
one 15 cm and one 4 cm aluminum cell; both cells have a diameter of 
6.35 cm. The sidewall thickness of each  cell was 178 $\mu $m, 
and entrance and exit window thicknesses were 71 $\mu $m and  
102 $\mu$m, respectively. As this experiment required only a liquid 
hydrogen target, only loop 3 was used; the other two 
loops were filled with helium gas at 0.12 MPa to save cooling power. 
The nominal temperature and pressure for the liquid hydrogen target during 
the experiment were 19 K and 0.17 MPa, respectively. The target density 
decreased by about 5 \% \cite{riad} at an incident beam current of 120 $\mu$A compared to 
its density at 10 $\mu$A (measured in an earlier experiment). 
The target assembly was housed inside a scattering chamber.

The scattering chamber in Hall A is divided into three sections. 
The vacuum in all three sections is maintained at a level of 0.13 mPa. 
The bottom section is 
fixed to the Hall A pivot and the top part of the scattering chamber contains 
the target's cryogenic plumbing. The middle section of the chamber
has an inner diameter of 103.7 cm, a wall thickness of 5 cm of aluminum and a height of 91 cm. 
The entrance and exit beam 
pipes are connected to this section.  The scattered particles go through
aluminum exit windows 18 cm high and 406 $\mu$m thick to the entrances of 
two high resolution spectrometers (HRSs). 

In order to reduce the heat deposition in a very small area of the target from 
an intense electron beam and to minimize corresponding target density changes,
the beam was rastered before it strikes the target. The fast 
rastering system is located 23 m upstream of the target. The rastering 
system contains two sets of magnets, one to deflect the beam vertically, and 
the other to deflect it horizontally. The magnetic field varies
sinusoidally at 17.7 kHz in the vertical direction and 25.3 kHz in the 
horizontal direction. The typical rastered beam spot size at the target was 
$\approx$ 3.5$\times$3.5 mm$^2$. 

A box located at the entrance window of each spectrometer can contain three movable 
collimators. The upper collimator is a stainless steel 5 mm thick sieve slit; it 
is used to study the optics of the spectrometers. The middle collimator is made of 
tungsten, and is  8 cm thick, 
6.29 cm wide, and 12.18 cm high, and is located at a distance of 110.9 cm from the target. 
The bottom position is empty and performs no collimation and is the one used in this experiment; 
the collimation is then defined by the aperture of the magnetic elements of the HRS.
The space between the exit window of the scattering chamber and the entrance window of 
each HRS consists of 20 cm of air. 

Elastic $ep$ events were selected by detecting scattered electrons and 
the recoiling protons in coincidence, using the two identical HRSs 
of Hall A. Each spectrometer consists of three 
quadrupoles and one dipole. The configuration 
is QQD$_n$Q, two quadrupoles followed by an indexed dipole ($n$=-1.25), and a quadrupole. Both 
spectrometers are designed for point-to-point focusing 
in the dispersive, and mixed focusing in the non-dispersive direction. The front 
quadrupole Q1 has a magnetic length of 0.941 m and an inner radius of 0.15 m,
and it is focusing in the dispersive direction. The quadrupoles 
Q2 and Q3 provide focusing in the non-dispersive direction and they 
both have an inner radius of 0.30 m and a magnetic length of 1.82 m. 
All three are superconducting, cos(2$\phi$) quadrupoles with an outside 
cylindrical magnetic field return iron yoke. The quadrupole fields are  
monitored using Hall probes and Gauss-meters.

The dipole in the HRS has a superconducting coil and warm iron configuration with shaped pole-faces and 
field gradient to help focus in the dispersive direction; 
its magnetic field deflects the  particles in a vertical 
plane by 45$^{\circ}$. The bending radius of the dipole is 8.40 m with a central gap 
of 0.25 m and an effective length of 6.6 m. The nominal maximum central trajectory 
momentum is 4~GeV/c, the momentum resolution is of the order of 10$^{-4}$, and the 
momentum acceptance is $\pm$ 5 \%. The field in the dipole is monitored and 
regulated by an NMR probe; it is stable at the 10$^{-5}$ level.

\begin{figure}
\begin{center}
\epsfig{file=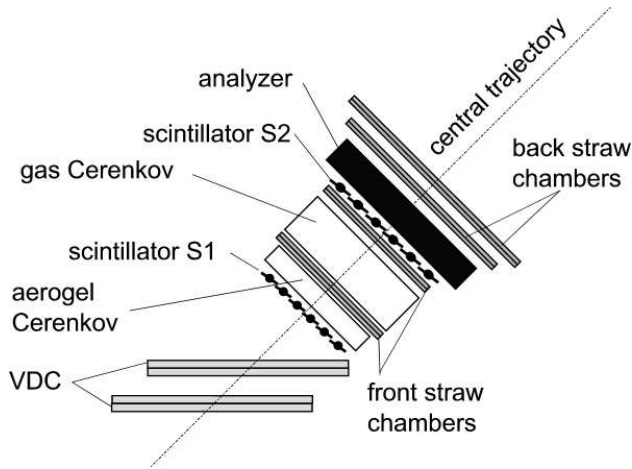, width=3.25in}
\vspace{0.1in}
\caption{Schematic of the hadron arm detector package including the polarimeter.}
\label{fig:haddet}
\end{center}
\end{figure}

The focal plane detector assembly for each spectrometer is enclosed in a metal 
and concrete shield house to reduce the  
background radiation. Both detector systems contain two vertical drift 
chambers (VDC), and scintillator arrays called S1 and S2. In addition, 
the electron detector system contains a gas \v{C}erenkov detector, and 
lead-glass arrays used as pre-shower and shower detectors; the 
hadron detector package contains an aerogel  \v{C}erenkov, 
a gas  \v{C}erenkov 
and the focal plane polarimeter (FPP). The assembly of the hadron 
arm detectors is shown in Fig. \ref{fig:haddet}. In this experiment 
only the VDCs and S1 and S2 detectors were used on the electron side, and the
VDCs, S1, S2 and the FPP on the hadron side. 

The two VDCs, installed close to the focal plane 
of each HRS, give precise reconstruction of positions and angles.
The central ray of the spectrometer passes through the 
center of each VDC at 45$^{\circ}$ to the vertical. The two VDCs 
are separated by 33.5 cm. 
The active area of each one is 211.8$\times$28.8 cm$^2$ with 
two wire planes at 45$^{\circ}$ to the dispersive 
direction and perpendicular to each other. All VDCs are operated at 
a high voltage of 4 kV, and the gas mixture used in these chambers 
is 62 \% argon and 38 \% ethane. The position resolution of 
each plane is $\approx$ 100 $\mu$m. 

In both HRS detector systems there are six scintillator paddles in plane 
S1 and six in S2. Each paddle is seen by one photomultiplier at each end.
The paddles in both planes are oriented such 
that they are perpendicular to the spectrometer central ray. The 
distance between the two planes is 1.933 m on the electron side and 
1.854 m on the hadron side. The active area of the S1 plane is 
170 $\times$ 36 cm$^2$, and of the S2 plane 212$\times$60 cm$^2$. The thickness of 
each paddle in both planes is 0.5 cm. The scintillator 
paddles of both planes overlap by 0.5 cm to achieve 
complete coverage of the focal plane area.  

The trigger for both spectrometers is similar and is formed from a coincidence 
between the signal from two scintillator planes, S1 and S2. The first 
requirement is to form a coincidence between the left and right signals from 
each paddle in the S1 and S2 planes. The time resolution per plane is about 
0.3 ns (1$\sigma$). These coincidence signals are fed into a memory lookup 
unit (MLU) which is programmed to form a second trigger, called ``S-Ray'' 
(trigger T1 for electron HRS and trigger T3 for hadron HRS), by requiring 
that the paddles that fired in the S1- and S2 planes belong to an allowed 
hit pattern. The allowed hit pattern requires that if paddle $\mid$N$\mid$ is 
fired in the S1 plane, then in the scintillator plane S2 a signal must come from 
paddle  $\mid$N-1$\mid$ or $\mid$N$\mid$ or $\mid$N+1$\mid$ or the overlap 
between two of those. Each spectrometer also 
has a ``loose'' trigger: for the electron arm it is 
formed by requiring that signals from two out of three detectors be present, 
S1-plane, S2- plane and the \v{C}erenkov detector; the hadron arm 
loose trigger requires signals from just S1-plane and S2-plane. 
These loose triggers are used 
to obtain detector efficiencies. Finally, the S-ray trigger signals 
T1 and T3 form the ``coincidence'' trigger T5 for the experiment. 
These five different triggers, T1 to T5, are sent to the trigger 
supervisor (TS) and are also counted by scalers. 

The TS was designed and built by the CEBAF online data acquisition (CODA) group. 
The functions of the TS include interface between the hardware trigger electronics 
and the computer data acquisition system, producing a computer busy signal that 
is used to calculate the computer dead-time, and pre-scaling of the trigger 
inputs T1 to T5.      
 
The data acquisition system was entirely developed at JLab by the CODA group\cite{coda}.
In this experiment CODA was running on a single Hewlett-Packard (HP9000) computer. 
The two main tasks of CODA are to transmit information from the detectors to the computer 
via read out controllers (ROC) and build events by collecting data from all the ROCs.  
The data are stored on a hard disk temporarily  for on-line analysis to monitor the experiment and
then transfered to tapes in the JLab mass storage system to be used later 
for final off-line analysis.

\subsection{Focal Plane Polarimeter}

Polarization experiments have become increasingly important in the 
study of nuclear reactions.  Focal plane polarimeters were 
standard equipment at intermediate energy proton accelerators, 
such as LAMPF \cite{lampf}, TRIUMF\cite{triumf}, SATURNE \cite{saturne}, 
and PSI \cite{psi}. Experiments at these facilities have demonstrated the 
sensitivity of spin observables to small amplitudes.
Similar considerations have more recently 
led to the development of proton polarimeters for use at electron 
accelerators, such as the MIT-Bates laboratory \cite{bates} and at 
the Mainz Microtron \cite{mainz}. 

The FPP in Hall A at JLab was designed, built, installed and calibrated
by a collaboration of Rutgers University, the College of William 
and Mary, Norfolk State University, the University of Georgia, 
and the University of Regina \cite{Mjones}. 

\subsubsection{Physical Description of Focal Plane Polarimeter}

The FPP is a part of the 
hadron detector package of the Hall A high resolution spectrometer. As shown in 
Fig. \ref{fig:haddet}, the polarimeter is installed downstream from the 
focal plane VDCs; it is oriented along the mean particle direction
in the focal plane area, at 45$^{\circ}$ to the vertical. It 
consists of two front detectors to track incident protons, followed by a carbon 
analyzer and two rear detectors to track scattered particles.

The four tracking detectors of the FPP are drift chambers made of straw tubes; 
the straw tube design is based on the one used for the EVA detector at 
Brookhaven National Laboratory \cite{eva}. 
The four drift chambers contain a total of 24 planes of straw tubes.
Twelve of these are in the 2 front chambers CH1 and CH2, where they are oriented along 
the $u$ and $v$ directions at $+ 45^{\circ}$
and $- 45^{\circ}$ relative to the spectrometer dispersive direction ($x$); each 
chamber  has the configuration $vvvuuu$. In the back chambers CH3 and CH4
the configuration is $uuvvxx$ and $uuuvvv$, respectively, where $x$ indicates 
that the $x$-coordinate is measured, and the straws are oriented along the $y$ direction. 

\begin{figure}
\begin{center}
\epsfig{file=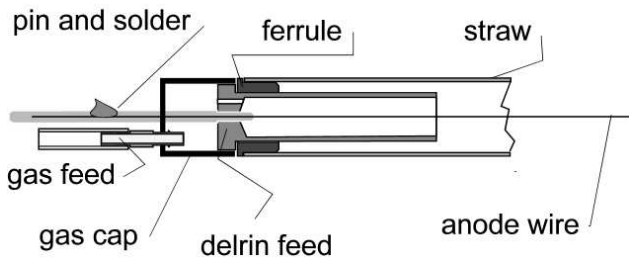, width=3.25in}
\vspace{0.1in}
\caption{Schematic showing the end assembly of straw tubes.}
\label{fig:strawend}
\end{center}
\end{figure}

The individual straws are thin walled mylar tubes containing the anode wire 
at their center, and the gas mixture. They are made by wrapping an inner 
10 $\mu $m thick aluminum  foil, and two 50 $\mu $m thick mylar foils 
around a mandrel, together with a heat setting glue. The inner diameter 
of the straw is 10.44 mm. The ground connection to outside is made with silver 
epoxy to a brass ferrule inserted at both ends. As shown in Fig. 
\ref{fig:strawend}, a delrin feed-through inserted
in the ferrule provides gas feed and exhaust, and a positioning hole 
for a brass slit pin, into which the high voltage anode wire is 
soldered under prescribed stress to insure that the gravitational 
sagging and electrostatic deflection of the wire are small. The anode wire is gold plated  
25 $\mu $m diameter tungsten-rhenium wire. 

The two front chambers are identical to one another and contain 1008 straw tubes each.
In these two chambers the straw tubes are precisely spaced by inserting 
their ends into aluminum blocks in which holes of diameter 10.75 mm have 
been drilled with 10.95 mm spacing center-to-center. Each block has 3 
layers of such holes separated vertically by 9.5 mm and shifted
by half a hole separation, providing a very tight packing. Each block 
accommodates 16 straws in each of the 3 layers, for a total of 48. The 
spacing between the straws of each plane are maintained with mylar shims 
glued every 30 cm along the length. The active area for the front chambers is 
60$\times$209 cm$^2$. The nominal distance between the two front 
chambers is 120 cm center to center; the intervening space
was occupied during this experiment by the 100 cm gas \v{C}erenkov 
detector: although not used in this experiment, it contributed an additional
3~mrad to the multiple scattering at the lowest proton energy. 

The two back chambers contain a total of 3102 straws. Each chamber 
contains six planes of straws, with the successive layers of straws 
glued together using precision guiding plates and pins to insure accurate
positioning, and is protected by 0.36 mm thick carbon fiber panels at the top 
and the bottom. Both chambers are positioned on a 1.9 cm thick and 31.5 cm wide plastic 
honeycomb, aluminum faced composite, that also provides a mounting 
surface for gas, high voltage distribution, and readout boards.
Chambers CH3 and CH4 have active areas of  
124 $\times$272 cm$^2$ and 142 $\times$295 cm$^2$, respectively. 
The distance between these two chambers is fixed and equal to 38.0 cm, 
center to center.

The gas mixture used in the FPP chambers is 62 \%/38 \% argon/ethane 
by volume. The straw chambers are operated at a high voltage of 1875 V. 
The drift velocity of electrons for this
gas mixture and for this voltage is about 50 $\mu$m/ns over almost the entire volume of the tube.
The efficiency of an individual straw tube for single track 
events is greater than 97 \% after correction for the small gap 
between them.

The analyzer consists of five sets of graphite plates 
with different thicknesses. Each set is made of two halves 
that can be moved separately on left and right. The plates 
are beveled at an angle of 45$^\circ$ so that the two halves 
overlap when closed. The plates have thicknesses of 1.9 cm, 
3.8 cm, 7.6 cm, 15.2 cm and 22.9 cm and they are separated by $\sim$ 1.6 cm. 
The ability to vary the thickness of the analyzer is necessary to optimize the 
efficiency while maintaining the Coulomb multiple scattering 
angle within acceptable limits. The carbon thicknesses used in this experiment 
at different proton energies are given in Table I.

The main contribution to small angle multiple
Coulomb scattering originates in the analyzer, with additional 
contributions from the scattering in the S2 paddles, the straw tubes in the two front 
chambers and the air between them. Table I gives typical multiple 
scattering angles for the relevant kinematics of this experiment. As Coulomb scattering is 
largely spin independent, in first order it does not affect the polarimeter 
performance; however, multiple scattering smears out the nuclear scattering 
distribution, and will therefore result in a small dependence of analyzing 
power upon the analyzer thickness. Coulomb scattering results in a strong forward peak 
of protons which did not undergo nuclear scattering; these events are 
suppressed by requiring a minimum scattering angle 2 to 3 times larger than 
the multiple scattering rms angle; here this minimum angle was fixed at 47~mrad (2.7$^\circ$). 

\begin{table}[h]
\caption{Multiple scattering for the ten proton kinetic energies. ${T_p}_{inc.}$ is 
the incident proton kinetic energy, $C_{thick.}$ the analyzer thickness, $\vartheta^{rms}$  
the root mean square Coulomb scattering angle, in FPP chambers (fpp), in the analyzer (C) 
and added quadratically (total).}
\begin{ruledtabular}
\begin{tabular}{cccccc}
${T_p}_{inc.}$ &  $C_{thick.}$ & $\vartheta ^{rms}_{fpp}$ & $\vartheta ^{rms}_{C}$ & 
$\vartheta ^{rms}_{total}$ \\   
(GeV)  & (cm) & (mrad) & (mrad) & (mrad) \\  \hline
0.267 &  7.62 & 7.9 & 16. & 17.8 \\
0.426 &  22.86 & 5.9 & 20.8 & 21.6 \\
0.639 &  41.91 & 3.8 & 18.1 & 18.5 \\ 
0.799 &  41.91 & 3.1 & 16.1 & 16.4 \\
0.959 &  49.53 & 2.8 & 14.8 & 15.1 \\
1.014 &  49.53 & 2.7 & 13.1 & 13.4 \\
1.156 &  49.53 & 2.2 & 11.6 & 11.8 \\
1.333 &  49.53 & 2.1 & 10.4 & 10.6 \\
1.599 &  49.53 & 1.8 & 9.3 & 9.5  \\ 
1.865 &  49.53 & 1.6 & 8.2 & 8.4  \\
\end{tabular}
\end{ruledtabular}
\label{tab:msangle}
\end{table}

To reduce the cost of electronics, the signal output of the individual 
straws is multiplexed in sets of eight. With multiplexing the maximum
rate each tube can accept safely is 100 kHz. One end of each straw 
is connected to the high-voltage distribution board, and the other to a readout 
board. The Rutgers University electronics shop
designed and built readout boards 
with 16 parallel channels. The input 
signal from each straw (typically 10 mV in size) is coupled to ground 
through a 1500 pF capacitor and fed into the input of an NEC1663 
amplifier. The amplifier output, a 100 mV positive signal, is fed 
into a LeCroy MVL407 quad comparator. This is a leading edge 
discriminator that gives a logical true when the input signal exceeds 
a supplied positive threshold voltage. The output of the comparator 
is then fed into pulse shaping circuitry. The readout board is 
divided into two halves, each of eight channels. The shaping 
circuitry for the eight channels gives a different width logic 
pulse for each of the channels. This allows all eight to be 
multiplexed together with OR chips into a single output channel, 
reducing the number of cables and channels of TDC needed. 
To limit potential noise problems, ground planes are inserted
within chamber readout card stacks, and differential output 
signals of amplitude 0.1 V are generated.
Level shifter boards located away from the chambers, near the TDCs, 
convert these signals to usual ECL levels for input to the TDCs.

The output pulse widths for eight channels, generally adjusted to 1-2 \% of the width, 
are given in Table II.
The identification of a wire within a group of 8 is obtained by decoding the
information from pipeline TDC's which digitize both the leading and trailing 
edge times of the signal. The wire group is given by the multiplexed output 
carrying a signal; the actual position of the track requires decoding
of the timing information to first identify the wire in the group and then
calculate the drift time using drift velocity calibration data. 
Multiple tracks within the same subset of 8 wires cannot be decoded, unless the hits are 
separated by at least 250 ns.

\begin{table}
\caption[]{The output pulse widths for the eight channels of the multiplexing circuitry 
on the chamber read out cards.}
\begin{ruledtabular}
\begin{tabular}{c|cccccccc}
Channel& 1 & 2 & 3 & 4 & 5 & 6 & 7 & 8               \\  \hline
Width (ns)& 25& 45 & 35 & 55 & 85-90 & 65 & 100-105 & 75  \\  
\end{tabular}
\end{ruledtabular}
\label{tab:pulse}
\end{table}

\section{DATA ANALYSIS}

The data were analyzed with the standard Hall A analysis program called ESPACE \cite{espace}. 
The output from ESPACE includes histograms, two-dimensional plots 
and multi-dimensional arrays (ntuples), which are used in further analysis 
to obtain quantities of interest, the $G_{Ep}/G_{Mp}$ ratio and the analyzing power $A_y$.

The kinematic settings of this experiment, the cuts applied in ESPACE, 
the selection of elastic events, the reconstruction of tracks in the FPP chambers, and 
the cuts on FPP variables are described in subsection A. A description  
of the azimuthal event distribution and asymmetry after scattering in the carbon analyzer, 
and spin transport through magnetic elements of the hadron HRS, is given in subsection B. 
The methods to calculate the $G_{Ep}/G_{Mp}$ 
ratio are described in subsection C, and subsection D describes the FPP calibration.

\subsection{Kinematic Settings and Selection of Good Events}

This experiment was performed at ten different values of $Q^2$; these $Q^2$ values 
as well as other useful kinematical quantities are given in Table III.

The Hall A analysis program ESPACE calculates the position 
and angle at the focal plane, $x^{fp},y^{fp},\theta^{fp},\phi^{fp}$, for each 
event from the raw VDC data.  The position, angles and 
momentum for proton and electron at the target, $y,\theta,\phi,\delta$, are then 
calculated using the HRS optics matrix; $\delta$ is the relative momentum 
$\delta$=$\frac{p-p_c}{p_c}$, 
with $p$ and $p_c$ being the scattered particle's momentum and the central momentum of the spectrometer, 
respectively.

\begin{table}[hbt]
\caption{Beam energies and spectrometer settings of the experiment. $E_{beam}$ is the beam energy, 
$Q^2$ is the four momentum transferred square, ${p_e}_c$, ${\theta_e}_c$, ${p_{p}}_c$ and 
${\theta_p}_c$ are central values of momentum and angle for the spectrometers detecting  
electrons and protons, respectively.} 
\begin{ruledtabular}
\begin{tabular}{cccccc}
$E_{beam}$ & $Q^2$ & ${p_e}_c$ & ${\theta_e}_c$ & ${p_p}_c$  & ${\theta_p}_c$  \\
(GeV) & (GeV$^2$)  & (GeV/c) &  (deg) &  (GeV/c) &  (deg)  \\ \hline  
0.934 & 0.50 & 0.675 & 52.59 & 0.756 & 45.28 \\ 
0.934 & 0.80 & 0.516 & 79.81 & 0.991 & 30.84 \\ 
1.821 & 1.20 & 1.193 & 43.45 & 1.268 & 40.36 \\ 
3.592 & 1.50 & 2.815 & 22.11 & 1.463 & 46.52 \\ 
3.592 & 1.80 & 2.656 & 25.01 & 1.649 & 42.92 \\ 
4.087 & 1.90 & 3.093 & 22.28 & 1.712 & 43.35 \\ 
4.087 & 2.17 & 2.950 & 24.40 & 1.872 & 40.68 \\
4.091 & 2.50 & 2.774 & 27.08 & 2.068 & 37.68 \\
4.091 & 3.00 & 2.507 & 31.29 & 2.357 & 33.59 \\ 
4.087 & 3.50 & 2.241 & 35.90 & 2.642 & 29.87 \\
\end{tabular}
\end{ruledtabular}
\label{tab:kine}
\end{table}
\noindent

\begin{figure}
\begin{center}
\epsfig{file= 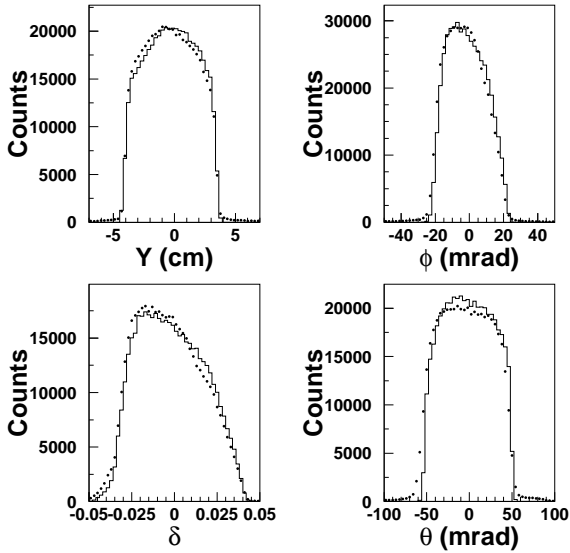,width=3.25in}
\caption{Comparison between the target variables $ y,\phi, \delta, \theta $ obtained from the data 
(dots) and calculated using the Monte Carlo program MCEEP (solid line) for protons detected in 
the right HRS. The MCEEP results are normalized 
by a factor of about 0.85,  agreement between the data and the MCEEP results is good.}
\label{fig:datamceep_p}
\end{center}
\end{figure}

\begin{figure}
\begin{center}
\epsfig{file= 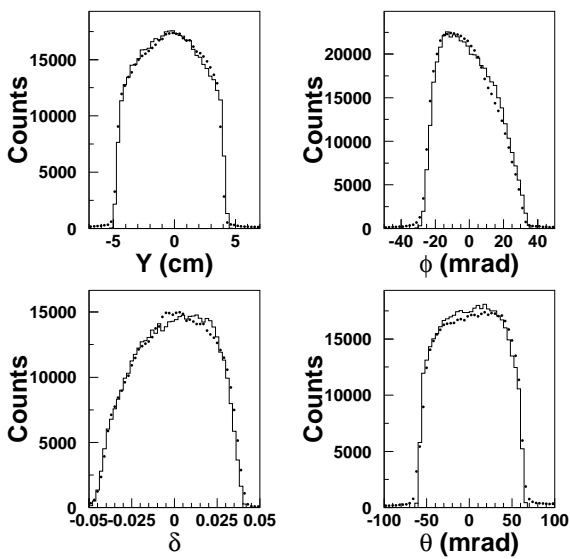,width=3.25in}
\caption{Comparison between the target variables $y,\phi, \delta, \theta$ obtained from the data (dots) 
and calculated using the Monte Carlo program MCEEP (solid line) for electrons detected in the left HRS. 
The MCEEP results are normalized by a factor of about 0.85, the agreement between the data and the 
MCEEP results is good.}
\label{fig:datamceep_e}
\end{center}
\end{figure}

The final optics matrix for each spectrometer was determined subsequent to this experiment using
the procedure as described in Ref. \cite{nilanga}.
Figs. \ref{fig:datamceep_p} and \ref{fig:datamceep_e} show 
a comparison between the distributions of the target variables $y,\phi, \delta, \theta$ obtained from the 
data and calculated using the Monte Carlo program MCEEP \cite{mceep}, for proton and 
electron at $Q^2$ of 3.5~GeV$^2$, respectively. The MCEEP results are normalized 
by a factor of about 0.85. This value seems quite reasonable, as in this experiment 
the trigger efficiency and the effect of boiling on target density were 
neither measured nor considered explicitly in the simulations. In addition, the BCM was not calibrated, 
as it would have been in the case of an absolute cross section measurement, for example. 
The agreement between the data and MCEEP results is good. Cuts were applied to all events 
for $\theta_{tar}$ ($\pm$65~mrad), $\phi_{tar}$ ($\pm$32~mrad), $y_{tar}$ ($\pm$6.5 cm) and 
$\delta$ ($\pm$5 \%), to eliminate the events seen in the tails of Figs. \ref{fig:datamceep_p} 
and \ref{fig:datamceep_e}.


The experimental event rates for each $Q^2$ and the one calculated with the Monte Carlo MCEEP
are given in the Table IV. The actual event rates are always lower than the MCEEP rates
by about 15 \%,  indicating that there is no significant background. The large 
difference seen between MCEEP 
and experimental event rate at $Q^2$ of 0.8~GeV$^2$ is due to a physical aperture cutting acceptance 
for the open collimator in the electron arm. Table IV also includes the total number of good events, 
average current, computer dead time, and average beam polarization value for each $Q^2$ value.

\begin{table}[hbt]                        
\caption{Experimental conditions.}
\begin{ruledtabular} 
\begin{tabular}{cccccccc}
$Q^2$ & Expt. & MCEEP & total no. & average & dead & average beam \\ 
      & rate  & rate  &of events  & current & time & polarization \\  
(GeV$^2$)& Hz & Hz & & ($\mu$A) & (\%) & \\ \hline
0.50 & 1050 & 1120 & 2.0 $\times$ 10$^6$  & 4 & 21 & 0.560$\pm0.030$ \\
0.80 & 250 & 420 & 4.6 $\times$ 10$^6$  & 10 & 1 & 0.544$\pm0.006$ \\
1.20 & 1100 & 1310 & 1.6 $\times$ 10$^7$  & 24 & 14 & 0.497$\pm0.012$ \\
1.50 & 420 & 480 & 9.2 $\times$ 10$^6$ & 10 & 4 & 0.483$\pm0.021$ \\
1.80 & 330 & 360 & 1.7 $\times$ 10$^7$  & 13 & 3 & 0.611$\pm0.020$ \\
1.90 & 1110 & 1160 & 5.6 $\times$ 10$^7$ & 63 & 24   & 0.391$\pm0.002$  \\
2.17 & 830 & 970 & 3.8 $\times$ 10$^7$ & 78 & 20  & 0.385$\pm0.002$  \\
2.50 & 540  & 650 & 5.6 $\times$ 10$^7$ & 74 & 4  & 0.390$\pm0.008$  \\
3.00 & 300  & 370 & 3.2 $\times$ 10$^7$ & 88 & 1  & 0.395$\pm0.002$  \\
3.50 & 140 & 160 & 2.0 $\times$ 10$^7$ & 94 & 1  & 0.384$\pm0.007$  \\
\end{tabular}
\end{ruledtabular}
\label{tab:rates}
\end{table}

\begin{figure}
\begin{center}
\epsfig{file=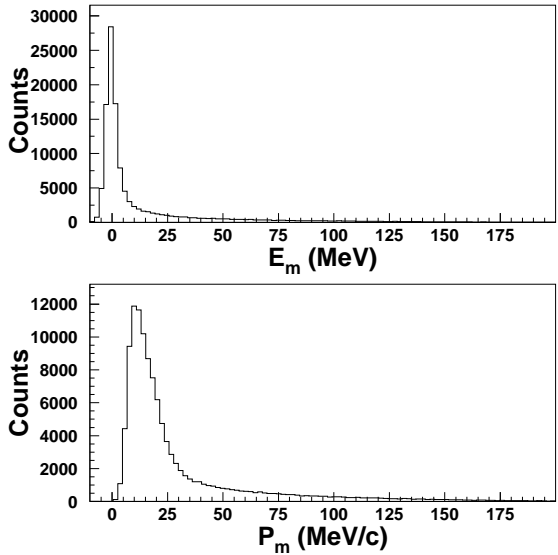, width=3.25in}
\vspace{0.1in}
\caption[]{Histogram of the missing energy and momentum $E_{m}$ and $P_{m}$, respectively. 
The peaks at $E_m = 0$ and at about $P_m$ = 10 MeV/c contain the elastic $ep$ events. The peak in $P_m$ 
has a larger width than $E_m$ because of finite angular resolution.}
\label{fig:empm}
\end{center}
\end{figure}

Selection of elastic $ep$ events was accomplished by implementing a correlated cut on the 
missing energy $E_{m}$ and missing momentum $p_m$. Due to the kinematic constraints 
of $ep$ elastic scattering, no further cuts were needed to remove background events. 
The missing energy $E_{m}$ is defined as:
\begin{equation}
E_m = E_{beam} + M_p - (E_{e} + E_{p}) 
\label{eq:em}
\end{equation}
\noindent
where $E_{p}$ is the scattered proton energy, and $M_p$ is the mass of 
the proton. From conservation of momentum, the missing momentum,
$P_{m}$, is defined as:
\begin{eqnarray}
P_m &=& \sqrt{ P_{m_{x}}^{2} + P_{m_{y}}^{2} + P_{m_{z}}^{2} } \label{eq:pm} \\
P_{m_{x}} &=& p_{e}\cdot \sin\theta _{e} - p_{p}\cdot\sin\theta_{p} 
\label{eq:pmt}\\
P_{m_{y}} &=& p_{e}\cdot \sin\phi_{e} - p_{p}\cdot\sin\phi_{p} 
\label{eq:pmn}\\
P_{m_{z}} &=& P_{beam} - (p_{e}\cdot \cos\theta _{e} + 
p_{p}\cdot\cos\theta_{p})\label{eq:pml}
\end{eqnarray}
\noindent
where $p_{e}$ ($p_p$) is the scattered electron (proton) momentum, 
$\theta _{e}$ ($\theta_p$) is the scattered electron (proton) Cartesian 
angle in the horizontal plane, and $\phi _{e}$  ($\phi_p$) is the 
scattered electron (proton) Cartesian angle in the vertical plane. 
In Fig.~\ref{fig:empm}, a histogram of the missing energy $E_m$ and missing momentum $P_m$
is shown. There is a peak at $E_m = 0$ and a peak at about $P_m = 10 MeV/c$. 
The peak in the $P_m$ histogram is not centered at zero because $P_m$ is defined positive, 
and it has a larger width due to finite angular resolution in both the transverse 
($\pm$2.0~mrad) and dispersive ($\pm$6.0~mrad) directions. A radiative 
tail is seen in the $E_m$ histogram up to about 125 MeV. The tail seen in the $P_m$ histogram 
includes the radiative tail as well as events that are multiple scattered in windows. 
 
In Hall A, two methods of  measuring the beam energy are now available, but at the time of 
this experiment neither was operational. The beam energy can be determined
using the $ep$ elastic kinematics, either from the measured scattering angles of the electron 
and proton or from Eq. (\ref{eq:em}) by forcing $E_m = 0$. Subsequent to this 
experiment, the beam energy was measured to a relative precision of about 
$1\times10^{-3}$ and the central momentum of the spectrometer was determined to 
the same precision; using this central momentum of the spectrometer and 
Eq. (\ref{eq:em}), the beam energies were determined and are listed in 
Table III. Comparison to the beam energies determined from 
the scattering angles lead to the conservative conclusion that the beam energy
was known with a relative precision of $2\times10^{-3}$.

Once an event was identified as an elastic $ep$ scattering, the next 
step was to search 
for a good track in the front and back FPP drift tube chambers. 
The analysis part for the FPP was incorporated in 
the main ESPACE program by the FPP group \cite{Mjones}; this part of the program 
reconstructs position and angles, in the front and back FPP chambers, then 
calculates the polar and azimuthal scattering angles, $\vartheta$ and  $\varphi$,
respectively, and the position of the interaction point, $Z_{fpp}$, in the carbon analyzer. 
Tracking in the chambers is done in the $u$ and $v$ coordinates separately.
Tracking starts with identifying 
clusters of hits in chambers CH1, CH2, CH3 and CH4. To determine a track for the
front (back) chambers, there must be at least one hit in CH1 and CH2 
(CH3 and CH4) and at least three hits total in the front (back) chambers. The 
efficiency for an individual straw to detect a proton is about 97 \%;
the total number of hits in the front chambers is usually about 5 or 6, and it is 4 or 5 for
the rear chambers (as the X-plane is not used). The total number of possible front (back) tracks is the 
number of clusters in CH1 (CH3) times the number of clusters in CH2 (CH4).
The straws have cylindrical symmetry so from a single hit
it is impossible to distinguish if the proton passed to the left or right of the center wire
(left/right ambiguity). For each possible track, a least-squares 
straight line fit is done for all possible
combinations of left/right for each hit. Then, out of all possible tracks,
the one with the smallest $\chi^2$ is selected as the good track. The only
exception is when the track with the smallest $\chi^2$ has only three hits;
then if there is another possible track with more hits, it is selected as the good 
track if its $\chi^2$ is reasonable.  

The FPP chambers are aligned by using a software procedure. The carbon 
can be moved out to make a clear path between the front and back FPP chambers. 
Then the tracks in the front and back chambers are aligned to the tracks in 
the VDC so that the FPP has the same coordinate system as the VDC. This 
is done for each FPP chamber by adjusting the three positions:
$z$ which is distance from the VDC and the zero of the $u$ and $v$ axes,
and adjusting the three angles of the FPP chamber (angle of the $uz$ plane, 
angle of the $vz$ plane and the angle of the $uv$ plane). The actual alignment
of the chambers was good, since in software the angles 
of the chambers are adjusted by less than a degree.

\begin{figure}[h]
\begin{center}
\epsfig{file=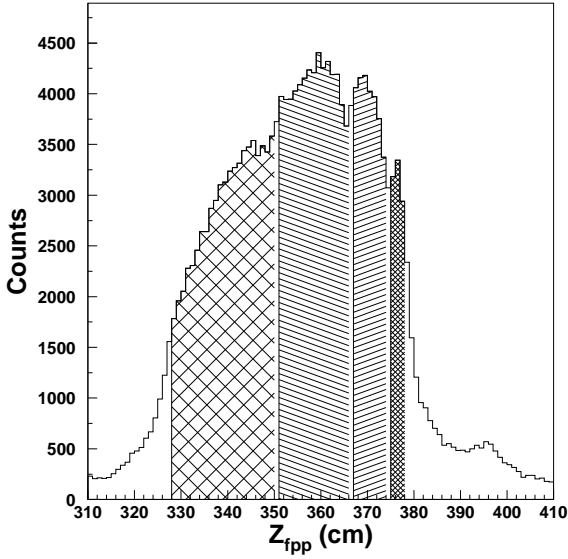,width=3.25in}
\vspace{0.1in}
\caption[]{Histogram of $Z_{fpp}$ for $Q^2$ = 3.5~GeV$^2$ with a total of 
49.5~cm of carbon consisting of 4 successive blocks of carbon with thickness of 
22.9, 15.2, 7.6 and 3.8~cm, indicated with different patterns.}
\label{fig:zfpp}
\end{center}
\end{figure}

A good track in the front and back FPP chambers is required to determine the
polar angle ($\vartheta$) and azimuthal angle ($\varphi$) of the scattered proton in the
carbon analyzer. Next, we calculate the distance of closest 
approach between the tracks from the front chambers and the back chambers, and at 
what distance, $Z_{fpp}$, from the VDC the closest approach occurred. 
In Fig.~\ref{fig:zfpp}, $Z_{fpp}$ is plotted for $Q^2$ = 3.5~GeV$^2$. The total thickness of carbon
is 49.5~cm; it consists of 4 successive blocks 
which are separated by $\sim$ 1.6 cm, so one can see a 
peak for each block in Fig.~\ref{fig:zfpp}. 
Chamber 3 is located at $Z_{fpp}$ of about 395~cm, as seen by the slight bump in
Fig.~\ref{fig:zfpp} from scattering in this chamber. For a selection of good events, a 
cut is placed on $Z_{fpp}$ depending on the thickness of carbon used for the 
given proton momentum. 

\begin{figure}[h]
\begin{center}
\epsfig{file=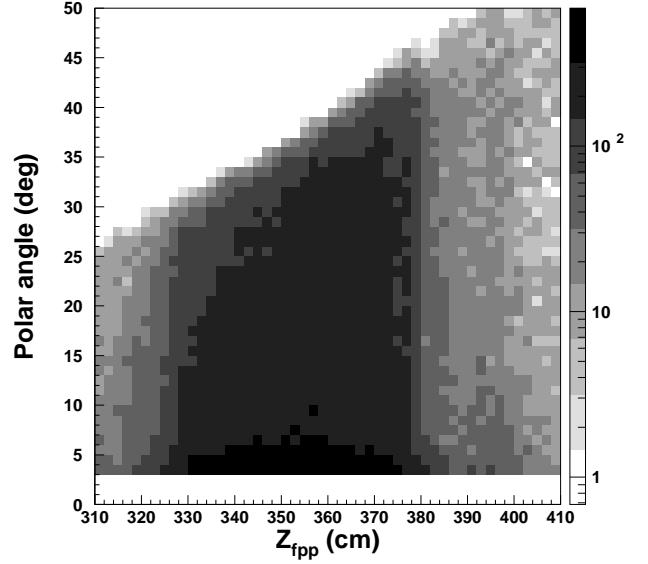,width=3.25in}
\vspace{0.1in} 
\caption[]{ Plot of $Z_{fpp}$ versus the polar angle $\vartheta$ with the
cone test applied.}
\label{fig:zfppthfpp}
\end{center}
\end{figure}

To reduce the false asymmetries a cone-test is applied for 
each event. An event passes
the cone-test when the track that hit the back chambers at the measured polar angle 
would have hit the chambers for any possible azimuthal angle. 
In Fig.~\ref{fig:zfppthfpp}, the polar angle ($\vartheta$) 
is plotted versus $Z_{fpp}$ with the cone-test applied. As expected, the 
range of accepted polar angles
increases as the interaction point in the carbon is closer to the back chambers.

\subsection{Azimuthal Asymmetry, Spin Transport in HRS and Extraction of $G_{Ep}/G_{Mp}$}

\subsubsection{Azimuthal Asymmetry}

Proton polarimeters are based on nuclear scattering
from an analyzer material like carbon or polyethylene; the proton-nucleus spin-orbit interaction 
results in an azimuthal asymmetry in the scattering distribution which can
be analyzed to obtain the proton polarization.
The azimuthal angular distribution of the yield, $I$, of 
scattered protons is: 

\begin{equation}
I~=~I_o~(1+A_y(\vartheta)\vec P^{fpp}\cdot \hat n),
\label{eq:genpol} 
\end{equation}

\noindent
where $I_o$ is the  unpolarized yield, $\vec P^{fpp}$ is the proton
polarization vector at the FPP, and 
$\hat n$ is a unit vector normal to the scattering plane defined as 
$\hat n =\hat k\times \hat k'/\mid \hat k\times \hat k'\mid$, 
with $\hat k$ and $\hat k'$, the unit vectors in the direction of the 
incident and scattered proton, respectively;  $A_y(\vartheta$) is the 
carbon analyzing power. 

\begin{figure}
\begin{center}
\epsfig{file=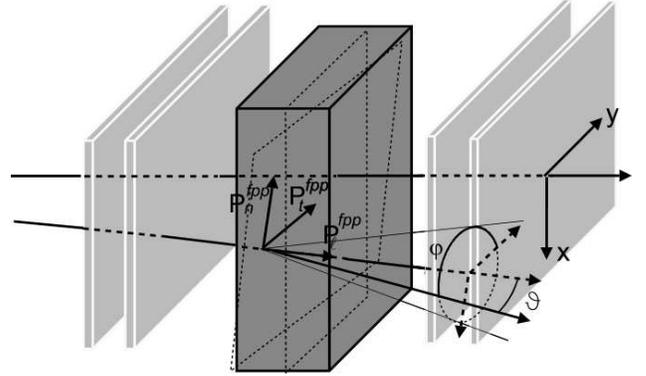,width=3.25in}
\vspace{0.1in}
\caption{Schematic of the polarimeter chambers and analyzer, showing  
a non-central trajectory; $\vartheta$ is the polar angle, and $\varphi$ is the azimuthal angle from 
the $y$-direction counterclockwise.}
\label{fig:pola}
\end{center}
\end{figure}

Fig. \ref{fig:pola} shows a schematic of the polarimeter chambers and analyzer. It shows a 
non-central trajectory, where $\vartheta$ is the polar scattering angle, and $\varphi$ is 
the azimuthal scattering angle defined  relative to the transverse direction in the polarimeter 
coordinate system. 

The detection probability for a proton scattered by the analyzer
with polar angle $\varphi$ and azimuthal angle $\varphi$ is given by:

\begin{equation}
f^\pm(\vartheta,\varphi) = \frac{\epsilon(\vartheta,\varphi)}{2\pi} 
\left (1 \pm A_y(P_t^{fpp}\sin{\varphi} - P_n^{fpp}\cos{\varphi}) \right ) 
\label{eq:azimuth}
\end{equation}
\noindent
where $\pm$ refers to the sign of the beam helicity, $P_{t}^{fpp}$ and $P_{n}^{fpp}$ are 
transverse and normal polarization components in the reaction plane at the analyzer, 
respectively;  $P_{\ell}^{fpp}$ is not measured because it does not result in 
an asymmetry as seen in Eq. (\ref{eq:genpol}). In Eq. (\ref{eq:azimuth}) the helicity-independent 
polarization has been omitted
because the polarization from two-photon exchange contributions
to elastic scattering is expected to be negligible.  
Here $\epsilon(\vartheta,\varphi)$ is an instrumental asymmetry that describes
non-uniformities in detector response that might result from
misalignments of the FPP chambers or from inhomogeneities in
detector efficiency. The non-uniformity of $\epsilon(\vartheta,\varphi)$ depends upon
the population of events on the detectors, which in turn is determined by the choice of kinematics.
However, recognizing that the detector response is helicity
independent and that non-uniformities are limited to a
few percent, we may approximate Eq. (\ref{eq:azimuth}) by,

\begin{equation}
f^\pm(\vartheta,\varphi) \approx \frac{1}{2\pi} 
\left(\epsilon(\vartheta,\varphi) 
\pm A_y(P_t^{fpp}\sin{\varphi} - 
P_n^{fpp}\cos{\varphi}) \right)
\end{equation}
\noindent
and obtain
\begin{equation}
\epsilon(\vartheta,\varphi) \approx \pi
\left(f^+(\vartheta,\varphi) + f^-(\vartheta,\varphi) \right)
\end{equation}

The instrumental asymmetry can then be described well by the
Fourier expansion

\begin{eqnarray}
\epsilon(\vartheta,\varphi) &=& \alpha_0(\vartheta) + \alpha_1(\vartheta)\cos{\varphi}+ \alpha_2(\vartheta)\cos{2\varphi}\nonumber \\
&+& \beta_1(\vartheta)\sin{\varphi} + \beta_2(\vartheta)\sin{2\varphi} + ...
\end{eqnarray}

The angular dependence of the probability distribution $f^{\pm}$ 
is approximated by the normalized yields,

\begin{equation}
Y^{\pm}_i = \frac{1}{\Delta \varphi} \frac{N^{\pm}_i}{N_{in}^{\pm}\eta(\vartheta)}
\end{equation}
\noindent
where the index $i$ refers to a bin in $\varphi$, $\Delta \varphi$
is the width of the bin,
$N^{\pm}_i$ is the number of events in bin $i$,
$N_{in}^{\pm}$ is the number of protons with specified 
helicity incident upon the FPP, and $\eta(\vartheta)$ is the
differential efficiency of the analyzing reaction, defined as the ratio of the 
number of protons scattered at angle 
$\vartheta$ with a good track in rear chambers over the number of incident protons 
with an acceptable track in the front chambers.
It is also useful to define sum and difference histograms,
\begin{eqnarray}
D_i &=& (Y^{+}_i - Y^{-}_i)/2 \\
E_i &=& (Y^{+}_i + Y^{-}_i)/2 
\end{eqnarray}
\noindent
whose expectation values are given by,

\begin{eqnarray}
\langle D_i \rangle &=& \frac{1}{2 \pi}\left(A_y(P_t^{fpp}\sin{\varphi_i} - 
P_n^{fpp}\cos{\varphi_i}) \right ) \\
\langle E_i \rangle &=& \frac{\epsilon_i}{2\pi}
\end{eqnarray}
\noindent
to lowest order in $\epsilon$.
Thus, $E_i$ measures the efficiency while $D_i$ is sensitive to
the transverse and normal components of the polarization at the
FPP.

\begin{figure}[h]
\begin{center}
\epsfig{file=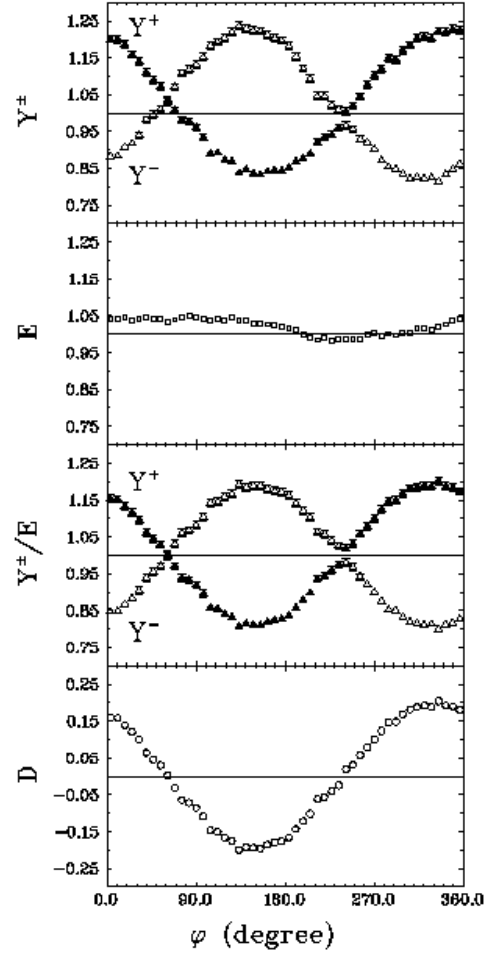,width=5in}
\vspace{-0.6in}
\caption{The azimuthal asymmetry distributions for $Q^2$=0.8~GeV$^2$. The first panel shows normalized yields
$Y^+$ and $Y^-$, the second panel shows the sum distribution $E$, the third panel 
shows normalized yield $Y^{\pm}/E$ and the fourth panel shows the difference asymmetry spectrum $D$.}
\label{fig:qsqr08}
\end{center}
\end{figure}

\begin{figure}[h]
\begin{center}
\epsfig{file=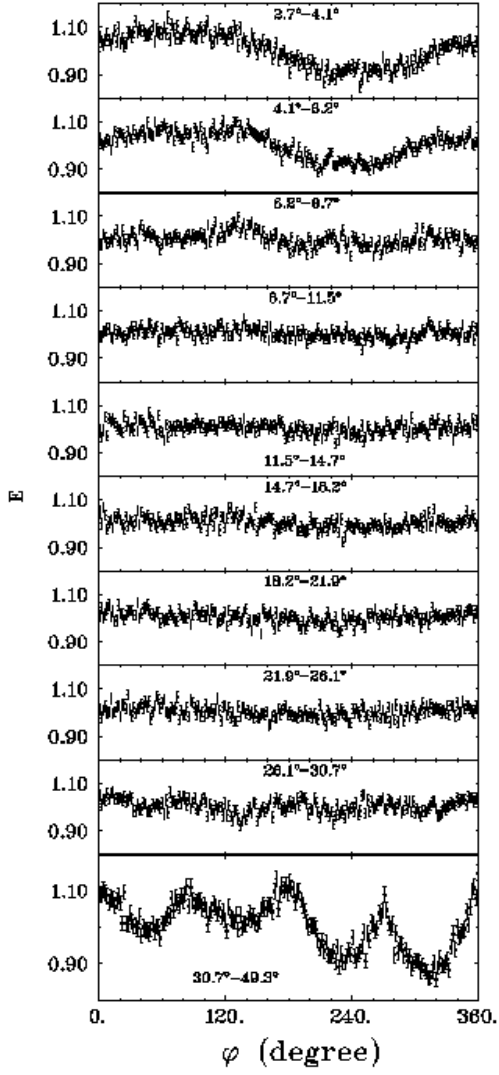,width=5in}
\vspace{-0.6in}
\caption{The sum $E=(Y^++Y^-)/2$ distributions for $Q^2$=0.8~GeV$^2$ for ten $\vartheta$ bins.}
\label{fig:qsq08}
\end{center}
\end{figure}

\begin{figure}[h]
\begin{center}
\epsfig{file=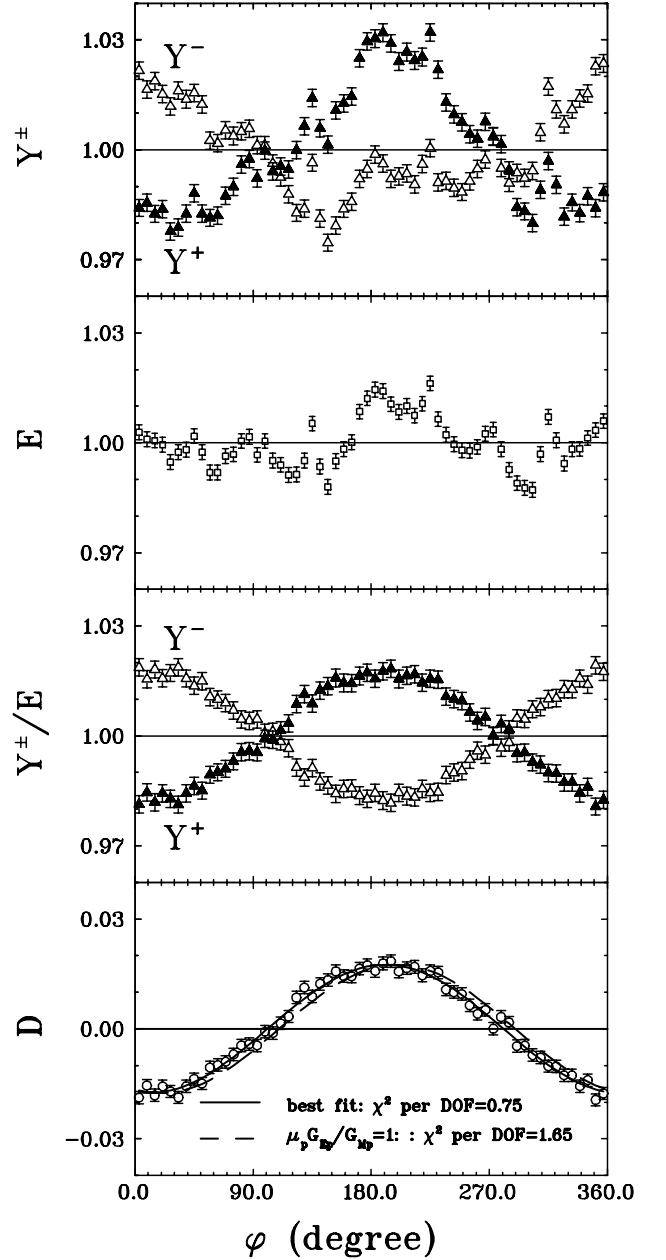,width=3.25in}
\vspace{0.1in}
\caption{The azimuthal asymmetry distributions for $Q^2$=3.5~GeV$^2$. The first panel shows normalized yields
$Y^+$ and $Y^-$, the second panel shows the sum distribution $E$, third panel 
shows normalized yield $Y^{\pm}/E$ and the fourth panel shows the difference asymmetry spectrum $D$ together with
sinusoidal fits and $\chi^{2}$ values for each curve.}
\label{fig:qsqr35}
\end{center}
\end{figure}

Figures \ref{fig:qsqr08} and \ref{fig:qsqr35} show the normalized yields and the sum and
difference spectra for $Q^2 = 0.8$ and 3.5 GeV$^2$, respectively.
Figures \ref{fig:qsq08} and \ref{fig:qsq35} show the efficiency spectra for the same
$Q^2$ settings binned with respect to polar angle.
Note that the widths of the scattering-angle bins were chosen 
to give approximately equal statistics.

The first panel in Figs. \ref{fig:qsqr08} and \ref{fig:qsqr35} shows helicity-dependent
normalized yields.
In an ideal polarimeter these distributions would exhibit
reflection symmetry about unity, but instrumental asymmetries
perturb that ideal pattern, especially for the larger
$Q^2$ where the physics asymmetry is smaller.
The second panel in each of these figures shows instrumental 
asymmetries at the level of several percent.
Figures \ref{fig:qsq08} and \ref{fig:qsq35} show these instrumental asymmetries 
divided into bins of the scattering angle $\vartheta$.
The smooth instrumental asymmetry seen in Fig. \ref{fig:qsq08} is clearly
dominated by the lowest two $\vartheta$ bins, which are sensitive
to chamber misalignment.
This effect is stronger for the lowest $Q^2$ data.
The experiment was performed in two periods separated by about
six weeks with the five lower $Q^2$ point taken in the second
period; no alignment data were taken at that time and the data
show that the chambers must have been slightly misaligned in
the intervening time. Nevertheless, the final results are not affected by this
misalignment because it is helicity independent and therefore cancels when the difference  
between the two helicity states is calculated.
The efficiency spectrum for $Q^2 = 3.5$ GeV$^2$ contains
higher frequency components arising primarily from 
large scattering angles, as shown in the bottom panel of
Fig. \ref{fig:qsq35}, that arise from edge effects.
These effects do not affect the final results either because
they are helicity independent also.
Furthermore, omission of the large $\vartheta$ bin simply
increases the statistical uncertainty in recoil polarization
by about 5 \% (relative) without altering the reported results.

The third panel in Figs. \ref{fig:qsqr08} and \ref{fig:qsqr35} shows normalized yields
corrected for instrumental asymmetry, namely $Y^{\pm}/E_i$.
These spectra show the reflection symmetry expected for
an ideal polarimeter. Finally, the bottom panel of these figures shows the
asymmetry spectra together with sinusoidal fits in Fig. 12 that can be described by:

\begin{figure}[h]
\begin{center}
\epsfig{file=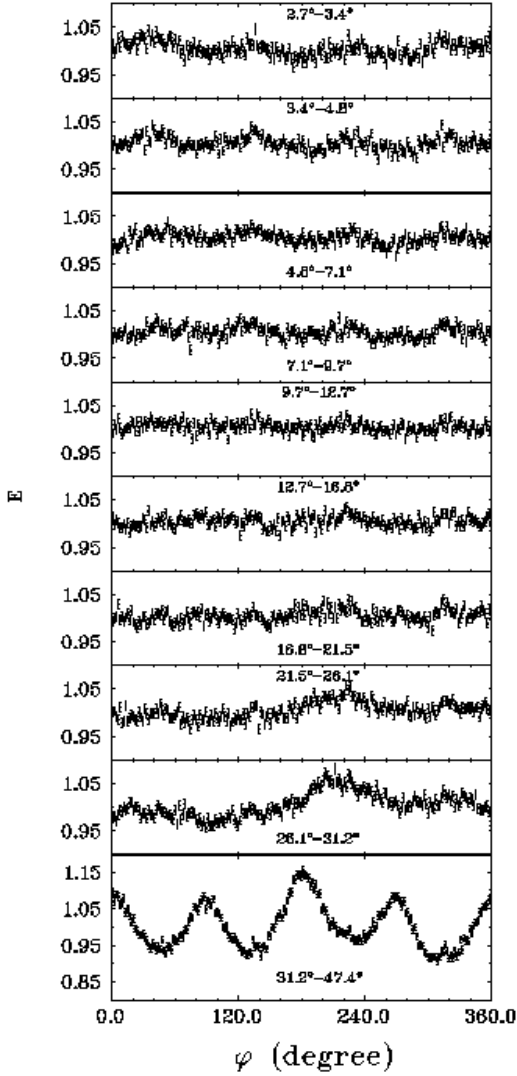,width=5in}
\vspace{-0.6in}
\caption{The sum $E_i=(Y^++Y^-)/2$ distributions for $Q^2$=3.5~GeV$^2$ for ten $\vartheta$ bins.}
\label{fig:qsq35}
\end{center}
\end{figure}

\begin{equation}
D_i = D_0 \cos{(\varphi + \delta)}
\end{equation}
\noindent
where the phase shift $\delta$ would be proportional to 
$\mu_p G_{Ep}/G_{Mp}$ in an ideal spectrometer.
For the final analysis spin precession in the spectrometer was
handled event-by-event using the procedures to be described
in subsequent sections.
For the present purposes it is sufficient to observe that
pure sinusoids fit the difference spectra with reduced
$\chi^2$ better than unity, showing that the instrumental 
asymmetries have been eliminated. 
Therefore, analysis of the difference spectrum is
insensitive to instrumental asymmetry.

\begin{figure}[h]
\begin{center}
\epsfig{file=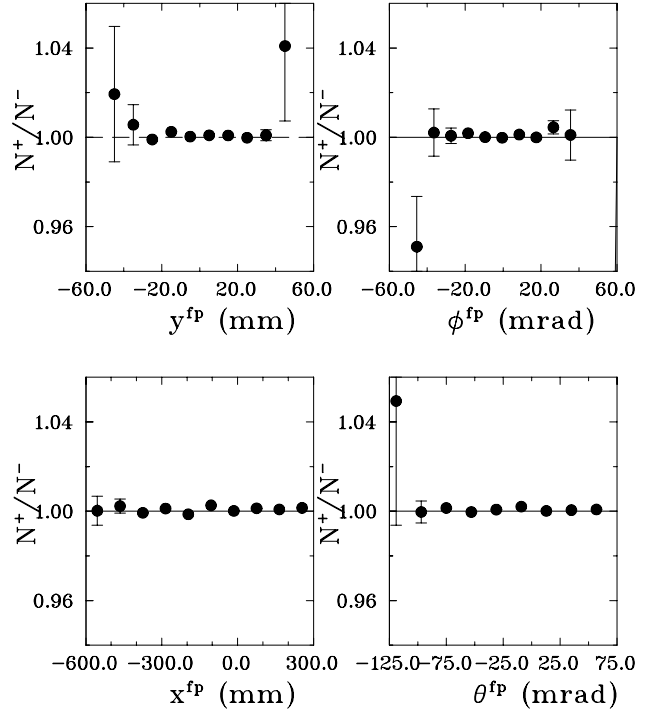,width=3.25in}
\vspace{0.1in}
\caption{The ratio $N^+$/$N^-$ at $Q^2$ of 3.5~GeV$^2$; the ratio is close 
to one and constant for each one of the four focal plane variables, 
thus indicating that there is no helicity 
dependence of $N^+$ and $N^-$ events in the focal plane.}
\label{fig:nplunminus}
\end{center}
\end{figure}

To demonstrate helicity independence at the focal plane, we show 
the ratio of helicity plus to minus events at $Q^2$ of 3.5~GeV$^2$ for variables 
$x^{fp}, y^{fp}, \theta^{fp}$, and $\phi^{fp}$ in Fig. \ref{fig:nplunminus}; the ratio is equal  
to one within statistical uncertainty (note the expanded y-scale) and is constant for 
each one of the four focal plane variables, thus indicating that there is no helicity dependence 
between $N^+$ and $N^-$ events in the focal plane of the hadron spectrometer.

\subsubsection{Spin Transport in HRS}

\begin{figure}[h]
\begin{center}
\epsfig{file=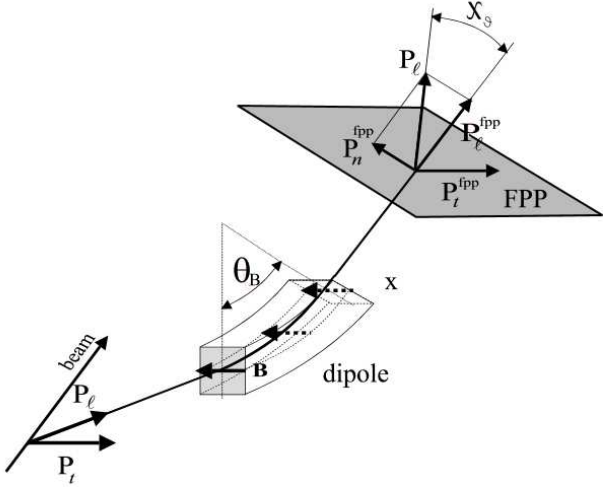,width=3.25in}
\vspace{0.1in}
\caption[]{Schematic drawing showing the precession by angle $\chi_{\theta}$ of the $P_{\ell}$ 
component 
of the polarization in the dipole of the HRS.}
\label{fig:precess}
\end{center}
\end{figure}

The proton spin precesses as it travels from the target to the focal plane
through the magnetic elements of the HRS as shown in Fig.~\ref{fig:precess}, 
and therefore the polarization components at the target and at the FPP are different. 
The hadron HRS in Hall A consists of 3 quadrupoles and one dipole with
shaped entrance and exit edges, as well as a radial field gradient.

The primary precession angle, $\chi _\theta$, in Fig. \ref{fig:precess} 
is defined as the difference 
between the spin and the trajectory rotation angles in the dispersive direction and can be expressed as: 

\begin{equation}
\chi _\theta=\gamma(\mu_p -1)(\theta_B + \theta -\theta ^{fp}),
\label{eq:chi}
\end{equation}

\noindent
where $\gamma$ is $E_p/M_p$, $\mu _p$ is the proton magnetic moment, 
($\theta_B + \theta -\theta ^{fp}$) is the bending angle 
in the dipole for a given event; $\theta^{fp}$ 
is the angle at focal plane, $\theta$ is the angle at the target, and
in the HRS $\theta_B$ = 45$^{\circ}$ for the central trajectory. 
As shown in Fig. \ref{fig:precess}, in first approximation for the homogeneous dipole field, 
the transverse polarization component $P_t$ is parallel to the field and does not precess; 
the longitudinal polarization component is perpendicular to the field and precesses with 
an angle $\chi _\theta$.  

For the HRS, the polarization vectors at the polarimeter, $\vec P^{fpp}$, are related to 
those at the target, $h\vec P$, where $\vec P$ is the polarization as given by Eqs.~
(\ref{eq:pn}), (\ref{eq:pt}), and (\ref{eq:pl}) through a 3-dimensional 
rotation matrix, $(\bf S)$, as follows:

\begin{center}

$\left( 
\begin{array}{c}
P_{n}^{fpp} \\ 
P_{t}^{fpp} \\ 
P_{\ell}^{fpp}
\end{array}
\right)
=
\left( 
\begin{array}{ccc}
S_{nn} & S_{nt} & S_{n\ell}  \\ 
S_{tn} & S_{tt} & S_{t\ell} \\ 
S_{\ell n} & S_{\ell t} & S_{\ell \ell}
\end{array}
\right) 
\left( 
\begin{array}{c}
P^{ind}_{n} \\ 
hP_{t} \\ 
hP_{\ell }
\end{array}
\right)$.
\end{center}
\noindent

where $h$ is the electron beam helicity, understood here to be the value of the longitudinal beam polarization 
component at the target, described with sign and magnitude.
Each one of the 9 matrix elements depends upon a particle's trajectory in the
spectrometer, and
therefore upon the 4 reconstructed kinematical variables 
$y,\phi ,\delta \mbox{ and }\theta$ of the recoil proton at the target. 
In the case of elastic $ep$ scattering, there is no helicity dependent polarization 
component $P_{n}$ at the target in the single-photon exchange approximation, however the helicity
independent component $P^{ind}_{n}$ may be induced due to the two-photon exchange mechanism, but 
this component is very small.   
The two polarization transfer components $P_{n}^{fpp}$ and  $P_{t}^{fpp}$ in
Eq. (\ref{eq:azimuth}) for each event, are then given by \cite{INPC98}:

\begin{eqnarray}
P_{n}^{fpp}&=&S_{nt}hP_{t} +S_{n\ell}hP_{\ell }~~~~\mbox{ and}
\nonumber \\
P_{t}^{fpp}&=&S_{tt}hP_{t} +S_{t\ell}hP_{\ell }.
\label{eq:pnpt}
\end{eqnarray}

With the assumption that there is only a dipole with an  
homogeneous field in front of the FPP, then $S_{tt}=1$, 
$S_{n\ell}=\sin\chi _\theta$, $S_{t\ell}=S_{nt}=0$, and Eq. (\ref{eq:pnpt}) 
simplifies to:

\begin{equation}
P_t^{fpp}=~hP_t \mbox{     and     } P_{n}^{fpp}=~hP_{\ell}~\sin \chi _\theta.
\label{eq:dipole} 
\end{equation}

However, the homogeneous dipole model is not appropriate for the HRS, as it
does not take into account the precession in the non-dispersive direction due to quadrupoles, 
fringe fields, and radial field gradient in the dipole.

A better approximation is to calculate the spin matrix elements $S_{ij}$ from the bend angles 
in the spectrometer, with several less restrictive assumptions
about the magnetic field configuration, (i) the longitudinal component
of the magnetic field with respect to the particle trajectory can be neglected, 
and (ii) the trajectory angles in the dipole change linearly along the path length
\cite{tnote}. The $S_{ij}$ 
elements under these assumptions are:

\begin{eqnarray}
S_{nt}&=-&\sin \chi_{\theta }~\sin \chi_{\phi }~\cos\chi_{\phi }' + O(\chi_{\phi} \times \chi_{\phi }' )\nonumber \\
S_{n\ell}&=&\sin \chi_{\theta }~\cos \chi_{\phi } \nonumber \\
S_{tt}&=&\cos \chi_{\phi }~\cos\chi_{\phi }' \nonumber \\
S_{t\ell}&=&\sin \chi_{\phi },
\label{eq:3dipole}  
\end{eqnarray}

\noindent
where the angles
\begin{eqnarray}
\chi _{\phi } &=& \gamma(\mu_p -1) [-(\phi -\phi ^{fp})+ (\phi ^d-\phi ^{fp})(1-\cos \chi_{\theta }) \nonumber \\
&-& \Delta\phi^d \left (1-\frac{2\sin\chi _{\theta }}{\chi _{\theta}}
+ \cos \chi_{\theta} \right ) ] \nonumber \\
\chi _{\phi }' &=& \gamma(\mu_p -1) [-(\phi^d -\phi^{fp})\sin \chi_{\theta } \nonumber \\
&+&\Delta\phi^d \left(\frac{2(1-\cos\chi _{\theta })}{\chi _{\theta}}- \sin\chi _{\theta} \right)]
\label{eq:chiphi}
\end{eqnarray}
represent the precession in the non-dispersive direction. 
Here $\phi -\phi ^{fp}$ is the corresponding trajectory total 
bending angle;
$\phi^d$ is the mean angle in the dipole,
and $\Delta\phi ^d$ is half of the bending angle
in the dipole in the non-dispersive plane.
In first order the later two angles
depend only on the non-dispersive target coordinate and angle:
\begin{eqnarray}
\phi ^d&=&(\phi ^d|y)y+(\phi ^d|\phi)\phi
\label{phid} \nonumber\\
\Delta \phi ^d&=&(\Delta \phi ^d|y)y+
(\Delta \phi ^d|\phi)\phi
\label{dphid}
\end{eqnarray}
The parameters, $(\phi ^d|y)$, $(\phi ^d|\phi)$, 
$(\Delta \phi ^d|y)$, and $(\Delta \phi ^d|\phi)$
are the couplings to the non-dispersive 
coordinate and angle at the target, and cannot be
measured directly; 
however they can be fitted and obtained from the data 
without using any spectrometer model. 
An analysis in \cite{tnote} also shows that for the HRS,
because of the relatively small angular acceptance and weak
fringe field effects, 
both being sources of the longitudinal field component,
assumption (i) above is fulfilled with good accuracy.
Comparison with the full COSY calculation described below, shows that assumption (ii) 
is also a good approximation (see  part IV C). 

The results presented here were obtained with spin matrix elements $S_{ij}$ calculated using a model 
of the HRS with quadrupoles, 
fringe fields, and radial field gradient in the dipole, for each tuning of 
the spectrometer and event by event with the differential-algebra-based 
transport code COSY \cite{bertz}. For a given central momentum, the 
output of the COSY code is a table of the expansion coefficients $C_{ij}^{klmnp}$ 
of the rotation matrix.
The matrix elements are calculated for each event using the $x$, $\theta$, $y$, $\phi$, and $\delta$ 
coordinates of the individual protons at the target with the form: 

\begin{equation}
S_{ij}= \sum_{k,l,m,n,p}C_{ij}^{klmnp} x^k \theta^{\ell} y^m \phi^n \delta^p.
\end{equation}
\noindent

The contribution to the systematic uncertainty due to the model parameters will be discussed in part IV C.

\subsection{Extraction of Ratio $G_{Ep}/G_{Mp}$} 

Results of this experiment published previously \cite{jones} were obtained using the analysis method 
of Ref. \cite{INPC98}, we will call this the ``old method''. The data presented here result from 
a complete re-analysis following a ``new method''. In Appendix A, both methods are described in some 
detail.

In both methods, the new and the old, the FPP data were analyzed in bins of polar scattering angle
$\vartheta$, such that statistics are approximately the same in all bins. The analyzing power
$A_y$ appearing in formulas in the Appendix A and in this section is an average analyzing power 
in each $\vartheta$ bin. The ratio $G_{Ep}/G_{Mp}$ was extracted for each bin and the ratio 
for a given $Q^2$ is the average over all $\vartheta$-bins.  

With the old method one calculates first the average asymmetries at the focal plane and
then transports them to the target using the average spin matrix elements, as discussed
in Appendix A 1. Assuming that the detector efficiency $\epsilon (\varphi )$ is a constant, and 
because the azimuthal scattering angle $\varphi$ in the analyzer is independent of the
target quantities $x$, $\theta $, $y$, $\phi $ and $\delta$, Eqs. (\ref{eq:meth1}) and 
(\ref{eq:methI}) from Appendix A 1 can be written simply as:

\begin{eqnarray}
\overline{A_{y} P_{n}^{fpp}}= hA_yP_{t}\overline{S_{nt}} + hA_yP_{\ell }\overline{S_{n\ell}}
\nonumber \\
\overline{A_{y}P_{t}^{fpp}}= hA_yP_{t}\overline{S_{tt}} + hA_yP_{\ell }\overline{S_{t\ell}},
\label{eq:pnptmean}
\end{eqnarray}
\noindent
where $\overline{{A_{y}P_{t}^{fpp}}}$  and $\overline{{A_{y}P_{n}^{fpp}}}$ are the 
effective asymmetries measured at the focal plane, and
$\overline{S_{ij}}$ are the mean values 
of the spin matrix elements; they are averages over the kinematical acceptance of the experiment. 
These linear equations must be inverted to obtain the target polarization asymmetries 
$hA_yP_{\ell}$ and $hA_yP_t$, which 
requires the determinant of the averaged spin transfer coefficients, shown below:

\begin{eqnarray}
\begin{array}{c} \left |
\begin{array}{cc}
{\overline{S_{nt}}} ~~
&\overline{S_{n\ell}} \\
\overline{S_{tt}} ~~
& \overline{S_{t\ell }} 
\label{eq:pnptmean1}
\end{array}
\right |.
\end{array}
\end{eqnarray}
 
When the precession angle $\chi _{\theta }$ is close to 180$^\circ$, the determinant is close to 
zero as the matrix element $S_{n\ell}~=~\sin \chi_{\theta }~\cos \chi_{\phi }$, and the spin 
transfer coefficients $S_{nt}$ and $S_{t\ell}$ are very small; hence the calculation 
of the polarization components at the target is less accurate; this occurs for $Q^2$ of 1.8, 
1.9, 2.5~GeV$^2$ and more so for $Q^2$ of 2.17~GeV$^2$.  

To overcome the difficulty associated with $\chi_{\theta}$ near $180^0$, 
a new method was developed to analyze the data; this 
method calculates the mean values of the asymmetries at the target rather than at the focal plane,
as described in Appendix A 2. Assuming, again for simplicity, 
that the efficiency $\epsilon (\varphi )$ is constant, the determinant for
the set of equations (\ref{eq:methII1}, \ref{eq:methII}) 
obtained with the new method in Appendix A 2, is:

\begin{eqnarray}
\begin{array}{c} \left |
\begin{array}{cc}
\overline{S_{nt}S_{n\ell }}+\overline{S_{tt}S_{t\ell }} 
&\overline{S_{n\ell}^2}+\overline{S_{t\ell }^2} \\
\overline{S_{nt}^2}+\overline{S_{tt}^2} 
& \overline{S_{nt}S_{n\ell }}+\overline{S_{tt}S_{t\ell }} 
\label{eq:pnptmean2}
\end{array}
\right |
\end{array}
\end{eqnarray}

\noindent

Now the determinant is non-zero even when 
$\overline{S_{n\ell}}$=0, because $\overline{S_{n\ell}^2}\neq$0,
even if $\overline{\chi _\theta}\simeq 180^0$.

Both methods give very similar results,
except for the region of $\chi _{\theta } \approx 180^0$,
where the old method gives larger uncertainties, and a slightly different result 
but still within the statistical uncertainty at the $Q^2$ of 1.8, 1.9 and 2.5~GeV$^2$.  
With the new method it became possible to obtain a ratio value at $Q^2$ = 2.17~GeV$^2$, which has  
$\overline{\chi _{\theta } }= 180^0$ for the central proton momentum. 

The values of the two asymmetries at the target, $hA_{y}P_{t}$ and 
$hA_{y}P_{\ell }$, obtained as discussed above, can now be used to calculate the ratio $G_{Ep}/G_{Mp}$ 
as shown below:

\begin{equation}
\frac{G_{Ep}}{G_{Mp}}=-\frac{hA_{y}P_{t}}{hA_{y}P_{\ell }}\frac{(E_{e}+E_{e^{\prime }})}{2M}\tan 
\frac{\theta _{e}}{2}
\label{eq:ratio1}
\end{equation}

As can be seen from this equation, neither the beam 
polarization nor the polarimeter analyzing power need to be known to extract the ratio, 
which results in very small systematic uncertainties.

\subsection{FPP Calibration}

Elastic $\vec{e}p\rightarrow e\vec{p}$ scattering provides a method to measure  
the analyzing power of the analyzer and therefore calibrate the FPP. 
After having obtained the ratio $G_{Ep}/G_{Mp}$ in a given kinematics following the 
method described in section III C, 
the values of the polarization observables at the target can be 
calculated from Eqs. (\ref{eq:pt}, \ref{eq:pl}) rewritten as:

\begin{center}
\begin{eqnarray}
P_t&=&-2\sqrt{\tau(1+\tau)}\frac{(G_{Ep}/G_{Mp})\tan \frac{\theta_e}{2}}
{(G_{Ep}^2/G_{Mp}^2+\tau/\epsilon)}\\
P_{\ell}&=&\frac{E_{beam}+E_e}{M_p}\sqrt{\tau(1+\tau)}\frac{\tan^2 \frac{\theta_e}{2}}
{(G_{Ep}^2/G_{Mp}^2+\tau/\epsilon)}
\end{eqnarray}
\end{center}

These are values of $P_t$ and $P_{\ell}$ at the target, averaged over the 
acceptance of the detectors. The asymmetry values $hA_yP_{t}$ and $hA_yP_{\ell}$ 
obtained from Eqs. (\ref{eq:methII1}, \ref{eq:methII}) in Appendix A 2, are then used to obtain
the analyzing power. For each proton energy, the average value of the beam helicity, $h$ 
from Table IV, was used to obtain the analyzing power $A_y$. 

\begin{figure}[h]
\begin{center}
\epsfig{file=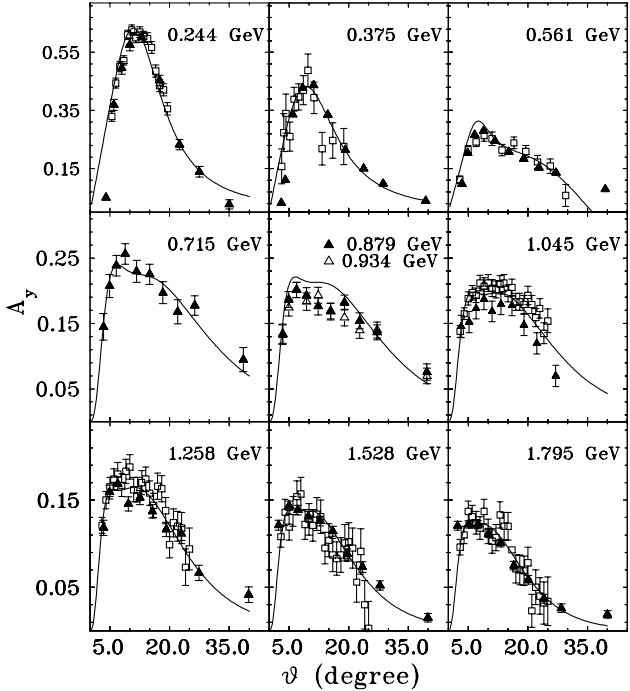,width=3.25in}
\vspace{0.1in}
\caption[]{Analyzing power as a function of polar scattering angle $\vartheta$ in FPP, 
for the JLab Hall A polarimeter (filled triangle and circle); also shown are fits as solid line and data
from other earlier calibration experiments as open squares.}
\label{fig:ay}
\end{center}
\end{figure}

The important properties of the FPP are the analyzing power $A_y(\vartheta)$ and 
the efficiency of the analyzing reaction, $\eta(\vartheta)$. 
The analyzing  power $A_y(\vartheta)$ is plotted as filled triangles in Fig. \ref{fig:ay} versus the polar 
scattering angle $\vartheta$ at the ten proton energies 
of this experiment; for proton energies 0.879 and 0.934~GeV the data have been placed in the same panel, 
as filled triangle and circle, respectively. 

In each panel the energy is that of the proton corrected for energy 
loss to half the thickness of the C-analyzer given in Table V. For 
comparison with the world data, fits adapted to the actual thickness of 
the analyzer used in this experiment are shown. At the two lowest proton 
energies the fits are from the calibration of Waters {\it et al.}  \cite{triumf}, 
and in the next two they are 
from the calibration of Ransome {\it et al.} \cite{lampf}. All 
others are from the 14 parameter fit of the Saclay calibration of Cheung {\it et al.} \cite{saturne}. 
In the same figure we also show as open squares data from various 
sources when available; in panel one the data at 0.225~GeV are from 
Ref. \cite{psi}, in panels two and three the data at 0.440 and 0.691~GeV are 
from Ref. \cite{lampf}, and in panels for 1.045~GeV to 1.795~GeV, the data are from 
Ref. \cite{saturne}. One must take notice that all earlier data were taken 
with significantly thinner C-analyzers. Overall the new data are in good 
agreement with fits and with previous data when available.

\begin{figure}[h]
\begin{center}
\epsfig{file=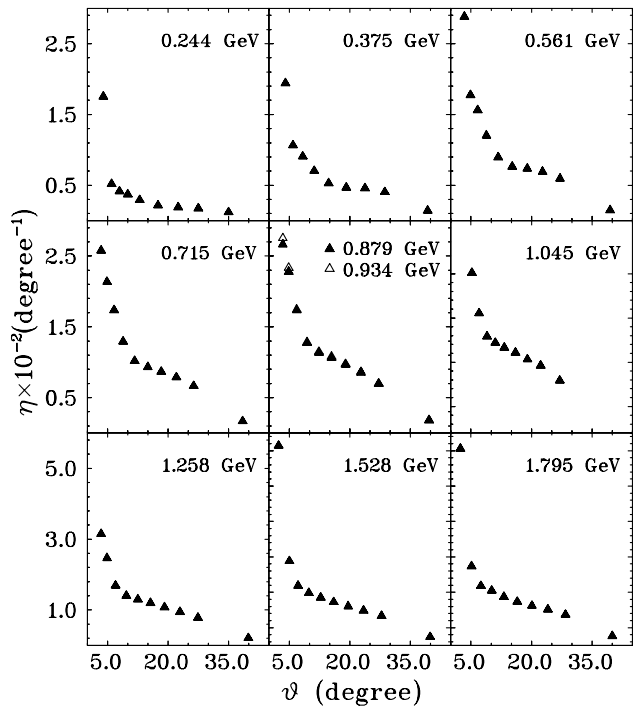,width=3.25in}
\vspace{0.1in}
\caption[]{Differential efficiency as a function of proton kinetic energy, ${T_p}_{analy.}$, evaluated at 
the half thickness of the analyzer.}
\label{fig:eff}
\end{center}
\end{figure}

The efficiency $\eta(\vartheta)$ for the ten values of 
$Q^2$ is shown in Fig. \ref{fig:eff}. For the 6 panels with the larger energies, the 
$\vartheta$ dependence is very much the same, as expected because the thickness of the C-analyzer 
was a constant 49.53 cm (see Table I). A cut was applied to eliminate all data below
2.7$^{\circ}$ and above 50.0$^{\circ}$. The lower cut was defined by the Coulomb scattering 
and the resolution of the drift chambers; the upper cut is determined by the size of 
the FPP back chambers. We include events
with smaller and with larger scattering angles than in Ref. \cite{saturne} because
we used drift chambers instead of the multi wire proportional chambers used in the calibration of 
Bonin {\it et al.} and Cheung {\it et al.}, and have therefore better position and angular resolution,
and also we used larger rear chambers. The enhancement seen 
at small angles is indicative 
of Coulomb scattering events from protons without nuclear interaction and with scattering 
angles larger than 2.7$^{\circ}$. Also as expected, the width of the forward peak 
is seen to decrease with increasing T$_p$.

Combining the $A_y(\vartheta)$ from Fig. \ref{fig:ay} and $\eta(\vartheta)$ of Fig. \ref{fig:eff} 
gives the differential figure of merit (FOM) $\eta(\vartheta)A_y^2(\vartheta)$. 
 The FOM values allow a quick 
evaluation of the number of ``good'' incident protons required to obtain a given statistical uncertainty 
in the polarization components. The calibration data for analyzing power $A_y$, efficiency $\eta$,
and FOM are given in Table V.

\begin{table}[h]
\caption[]{Averaged analyzing power, $\overline A_y$,  and integrated efficiency $\eta_{int.}$ and 
figure of merit, FOM, of the FPP for the ten proton kinetic energies of the experiment, ${T_p}_{analy.}$. 
The range of integration for all three quantities is  $2.7^{\circ} < \vartheta < 50^{\circ}$. 
Both the statistical (stat.) and systematic (sys.) uncertainties for ${FOM}$ are given.}
\begin{ruledtabular}
\begin{tabular}{ccccc}
${T_p}_{analy.}$ & $\overline A_y \pm stat$ & $ \eta_{int.} $ & ${FOM}\pm stat.\pm sys.$ \\  
(GeV) &  (cm) &  & $\times 10^{-2}$ \\  \hline
 0.244  & 0.274 $\pm$ 0.006 & 0.115 & 1.32  $\pm$  0.04 $\pm$ 0.08\\ 
 0.375  & 0.223 $\pm$ 0.002 & 0.244 & 1.57  $\pm$  0.02 $\pm$ 0.10\\ 
 0.561  & 0.183 $\pm$ 0.002 & 0.298 & 1.13  $\pm$  0.02 $\pm$ 0.07\\ 
 0.715  & 0.195 $\pm$ 0.005 & 0.323 & 1.25  $\pm$  0.05 $\pm$ 0.09\\ 
 0.879  & 0.162 $\pm$ 0.004 & 0.358 & 0.98  $\pm$  0.03 $\pm$ 0.09\\ 
 0.934  & 0.156 $\pm$ 0.004 & 0.362 & 0.94  $\pm$  0.03 $\pm$ 0.09\\ 
 1.045  & 0.145 $\pm$ 0.008 & 0.382 & 0.83  $\pm$  0.06 $\pm$ 0.05\\
 1.258  & 0.124 $\pm$ 0.003 & 0.401 & 0.65  $\pm$  0.02 $\pm$ 0.06\\ 
 1.528  & 0.099 $\pm$ 0.002 & 0.425 & 0.50  $\pm$  0.01 $\pm$ 0.04\\
 1.795  & 0.080 $\pm$ 0.002 & 0.443 & 0.35  $\pm$  0.01 $\pm$ 0.03\\
\end{tabular}
\end{ruledtabular}
\label{tab:aysqreff}
\end{table}

The statistical uncertainty in the results of this experiment is directly 
dependent on the analyzing power $A_y$ and the differential efficiency $\eta$, and can be written as:

\begin{equation}
\Delta P_n^{fpp}=\Delta P_t^{fpp}=\sqrt{\frac{2}{\eta A_y^2(N_{tot})}}=
\sqrt{\frac{2}{FOM(N_{tot})}}
\end{equation}
\noindent
where $N_{tot} = N_{in}^+~+~N_{in}^-$ is the total number of incident protons  
on the analyzer.

\section{RESULTS AND DISCUSSION}

\subsection{Experimental Results}

\begin{figure}[h]
\begin{center}
\epsfig{file=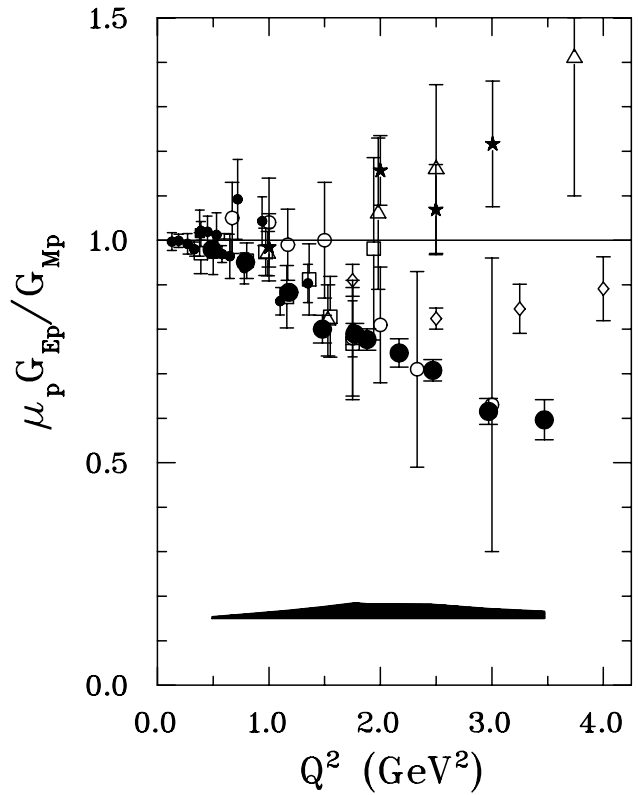,width=3.25in}
\vspace{0.1in}
\caption[]{The ratio $\mu_p G_{Ep}/G_{Mp}$ from this experiment, filled circles, compared
to world data, shown with the same symbols as in Fig. \ref{fig:gepgd_gmpgd}, 
with the 2 data points of Ref. \cite{milbrath} shown with filled squares.
The absolute value of systematic errors from this experiment
is shown  by the shaded polygon.}
\label{fig:gepgmp_prl}
\end{center}
\end{figure}

Figure~\ref{fig:gepgmp_prl} shows the results for the  ratio $\mu_p G_{Ep}/G_{Mp}$ from this experiment 
as filled circles, and from all experiments 
that took place after 1970 with same symbols as in Fig. \ref{fig:gepgd_gmpgd}. 
Our data points are shown with statistical 
uncertainties only; the systematic uncertainties are indicated by the 
black polygon. The data are tabulated in Table VI, where statistical and 
systematic uncertainties, as well as  $Q^2$ bin sizes are given for each data point. 
The results from this 
experiment differ from those previously published \cite{jones} in three ways: the points at 1.77, 
1.88  and 2.47~GeV$^2$ have moved slightly as a result of the new analysis method described in 
subsection III C and Appendix A, a new point was obtained at Q$^2$=2.13~GeV$^2$, and the systematic 
uncertainties are 2-3 times smaller. 

\begin{table}[hbt]
\caption{The ratio $\mu_p G_{Ep}/G_{Mp} \pm$ statistical uncertainty
(1$\sigma$). $\Delta_{sys}$ is the systematic uncertainty from 
Table VII. $\overline Q^2$ and $\overline \chi_{\theta}$ are the weighted 
average four momentum transfer squared and spin precession angle, respectively. $\Delta Q^2$ is
half the $Q^2$ acceptance. The last column $P_t/P_{\ell}$ is the ratio of measured polarization components 
at the target, the relative uncertainty is the same as for $\mu_p G_{Ep}/G_{Mp}$.}
\begin{ruledtabular}
\begin{tabular}{cccccc}
$\overline Q^2 \pm\Delta Q^2$ & $\overline \chi_{\theta} $ & $\mu_p G_{Ep}/G_{Mp}$  & $\Delta_{sys}$  & $P_t/P_{\ell}$  \\ 
(GeV$^2$) & (deg) & ($\pm$ stat. uncert.)  &     & \\ \hline
0.49$\pm$.04 &  105 & 0.979 $\pm$ 0.016 & 0.006 & -0.822  \\ 
0.79$\pm$.02 &  118 & 0.951 $\pm$ 0.012 & 0.010 & -0.527  \\ 
1.18$\pm$.07 &  136 & 0.883 $\pm$ 0.013 & 0.018 & -0.492  \\ 
1.48$\pm$.11 &  150 & 0.798 $\pm$ 0.029 & 0.026 & -0.422  \\ 
1.77$\pm$.12 &  164 & 0.789 $\pm$ 0.024 & 0.035 & -0.381  \\ 
1.88$\pm$.13 &  168 & 0.777 $\pm$ 0.024 & 0.033 & -0.368  \\
2.13$\pm$.15 &  181 & 0.747 $\pm$ 0.032 & 0.034 & -0.329  \\ 
2.47$\pm$.17 &  196 & 0.703 $\pm$ 0.023 & 0.033 & -0.284  \\ 
2.97$\pm$.20 &  218 & 0.615 $\pm$ 0.029 & 0.021 & -0.224  \\ 
3.47$\pm$.20 &  239 & 0.606 $\pm$ 0.042 & 0.014 & -0.198  \\
\end{tabular}
\end{ruledtabular}
\label{tab:results}
\end{table}

\begin{figure}[h]
\begin{center}
\epsfig{file=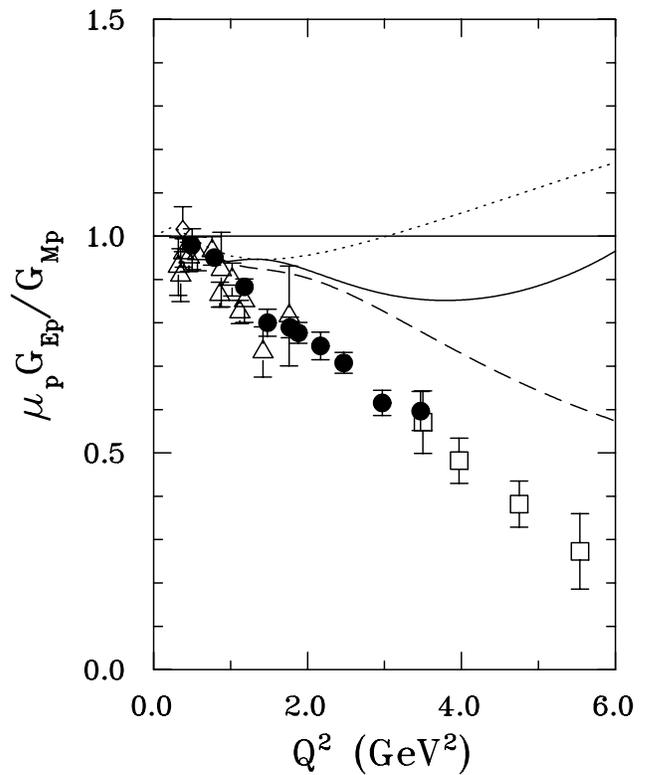,width=3.25in}
\vspace{0.1in}
\caption[]{The ratio $\mu_p G_{Ep}/G_{Mp}$ obtained in this experiment, and other 
polarization transfer
experiments, including~\cite{milbrath,pospischil,gayou,gayou2}. 
The curves are fits to the Rosenbluth separation data: one from 
Bosted~\cite{bosted2} shown as dotted line, and two from \cite{rosenblut} shown as 
solid line for all world data, and a dashed line for a selected subset of the world data. }
\label{fig:rosenfit}
\end{center}
\end{figure}
As seen in  Fig. \ref{fig:gepgmp_prl}, the electric form factor data from the past 30 
or so years can be described by the dipole form factor, $G_{Ep}/G_{D} \approx 1$ 
considering the spread in the data from different experiments; however, the data 
from this experiment deviate from the dipole form factor significantly 
starting at $Q^2$ of 1~GeV$^2$. Note that the results from this experiment are in apparent good agreement 
with the earlier data of Refs. \cite{berger,price,bartel} which have much larger 
uncertainties; also considering the larger uncertainties, the SLAC data of Ref. \cite{andivahis} 
are compatible with our results up to about $Q^2$ of 2.5~GeV$^2$.

Several global analyses of the $\mu_pG_{Ep}/G_{Mp}$ ratio obtained by 
Rosenbluth separation method have been done. We show in Fig.~\ref{fig:rosenfit}
three of these fits: the dotted line is the Bosted fit \cite{bosted2}. 
The solid  and dashed lines are from a recent 
global fit by Arrington~\cite{rosenblut}, including all existing data, and a 
subset of the data, respectively. The subset of the data is a biased selection 
which was chosen to give the lowest possible ratio and therefore represents 
a lower limit for the ratio extracted from the cross section data alone. 
All fits tend to be dominated by the more recent 
SLAC data. Only the $\mu_pG_{Ep}/G_{Mp}$ ratios obtained by recoil 
polarization are shown here; the data from this 
paper, from ref. \cite{gayou}, the early data of Milbrath et al. 
\cite{milbrath}, Pospischil et al. \cite{pospischil}, and a 
collection of measurements made in various Hall A experiments and analyzed 
by Gayou et al.\cite{gayou2}. We conclude from the comparison between 
Rosenbluth and polarization data shown in Fig. \ref{fig:rosenfit} and from the 
further analysis done in ref. \cite{rosenblut}, that the 
two methods give definitively 
different results; the difference cannot be bridged by either simple re-normalization 
of the Rosenbluth data or by variation of the polarization  data within the 
quoted statistical and systematic uncertainties.

The new results are the first high precision measurements of the ratio  $G_{Ep}/G_{Mp}$
above $Q^2$ of 1.0~GeV$^2$. The result for 
the ratio shows a systematic decrease 
as $Q^2$ increases from 0.5~GeV$^2$ to 3.5~GeV$^2$, which 
indicates that $G_{Ep}$ falls faster than $G_{Mp}$. In the non-relativistic 
limit, this fact could be interpreted as indicating that the spatial 
distributions of charge and magnetization currents in the proton are definitely different.

\subsection{Nature of the Data}

\begin{figure}[h]
\begin{center}
\epsfig{file=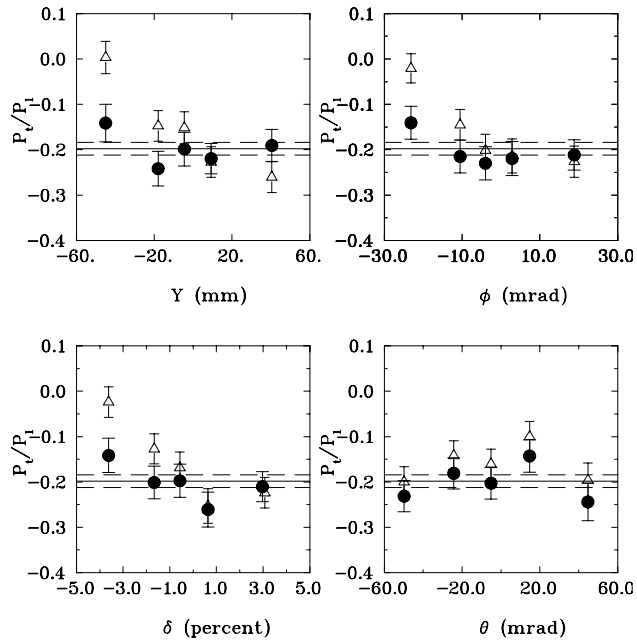,width=3.25in}
\vspace{0.1in}
\caption{The ratio $P_{t}/P_{\ell}$ at the target, plotted as a function of the 4 
kinematical target variables for Q$^2$=3.5~GeV$^2$. The filled circles are the results 
from COSY; the open triangles are the values obtained not taking 
into account the spin rotation due to the quadrupoles and the field gradient and fringe field 
of the dipole. Also shown is the range of statistical uncertainty corresponding to the uncertainty on 
$\mu_p G_{Ep}/G_{Mp}$.}
\label{fig:ptpltgt35}
\end{center}
\end{figure}

To illustrate the stability of the polarization data presented in the previous part, 
we show in Fig. \ref{fig:ptpltgt35} the variation of the the polarization 
component ratio, $P_{t}/P_{\ell}$ at the target, as a function of 
the four target variables: $y,~\phi,~\delta~$ and $\theta$, for $Q^2$=3.5 GeV$^2$. 
In each one of the panels, the data 
are integrated over the full experimental acceptance for the other three variables; 
therefore the statistics in each panel are not independent. The filled circles are the results 
obtained with full COSY calculation; the open triangles are the values obtained not taking 
into account the spin rotation due to the quadrupoles and the field gradient and fringe field 
of the dipole. The scatter in the data (filled circles) for each variable is compatible 
with the statistical uncertainty on $P_{t}/P_{\ell}$, 
which are shown as the dashed lines below and above the average, shown as the solid line; 
a constant value of $P_{t}/P_{\ell}$ over the acceptance of the spectrometer in each one of the four target 
variables demonstrates that the spin transport matrix elements are correct over the 
whole acceptance of the spectrometer, and indirectly, that the spectrometer model used in COSY is correct.
The values of $P_{t}/P_{\ell}$ are given in Table VI.

\begin{figure}[h]
\begin{center}
\epsfig{file=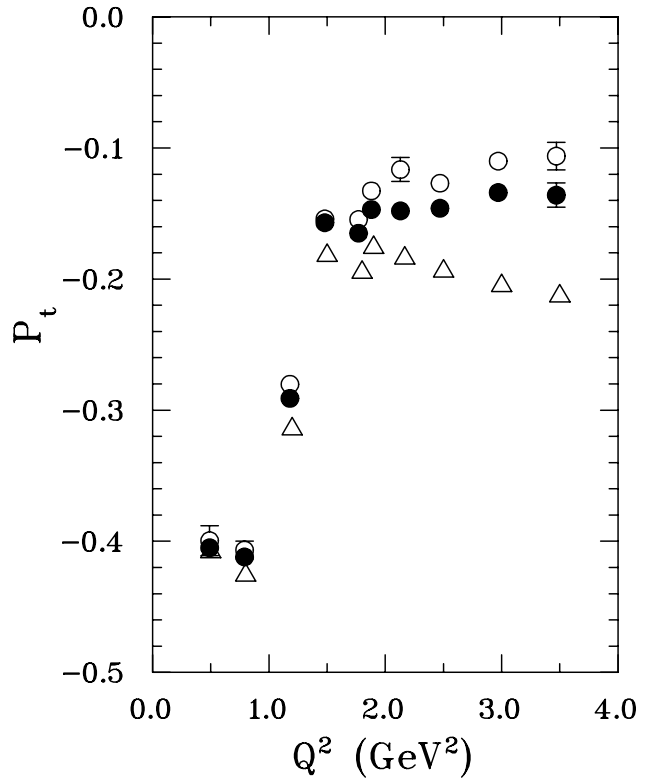,width=3.25in}
\vspace{0.1in}
\caption{The transverse component of the polarization at the target, $P_t$ measured in this experiment is 
shown as solid circles. The open triangles show the value of $P_t$
calculated using the dipole approximation, or scaling assumption: $G_{E}=G_{D}$ and 
$G_{M}=\mu_p G_{D}$. Also shown as open circles are the values of $P_t$  from the approximation 
$P_{t}=P_{t}^{fpp}$, not taking into account the spin rotation due to the 
quadrupoles and the field gradient and fringe field of the dipole.}
\label{fig:ptvsq2}
\end{center}
\end{figure}

The systematic decrease of the ratio $\mu_{p}G_{Ep}/G_{Mp}$ shown in Fig. \ref{fig:gepgmp_prl} 
and given in Table VI can be traced to the observed absolute values of $P_t$ being 
systematically and increasingly smaller than the values calculated using the dipole 
form factor $G_D=(1 + Q^2/0.71)^{-2}$, and $G_{Ep}/G_D$ and $ G_{Mp}/\mu_{p} G_D$. This is 
demonstrated in Fig. \ref{fig:ptvsq2}, where the values of $P_t$ at the target obtained in this experiment 
are shown versus Q$^2$ as solid circles. Also shown in this figure are the values of $P_t$
calculated from Eqs. (\ref{eq:pt}) and (\ref{eq:pl}) with dipole form factors as open triangles, 
and open circles are $P^{fpp}_t$ calculated from the data. It is seen 
from this figure that the additional polarization rotation 
introduced by the quadrupoles, and fringe and indexed fields in the dipole, is always small, and tend 
to decrease the magnitude of $P^{fpp}_t$ from its value at the target. 
The spin transport coefficient in Eq. \ref{eq:pnptmean} $\overline S_{t\ell}$
is small but not zero; it would average to zero over a symmetric angular distribution of events in  
the HRS acceptance. However, the $Q^2$ dependence 
of the cross section results in more events at small scattering angle than large, and that 
results in a non-zero average value of $S_{t\ell}$, producing the small change of 
the observed transverse component at the target seen in Fig. \ref{fig:ptvsq2}. 
Noticeable is the abrupt bend for all three sets; this apparent discontinuity 
is the result of the choice 
of beam energies: the last five points were taken at constant beam energy, 
whereas the first five where taken with an increasing energy to optimize 
the experiment. 

\subsection{Discussion of Systematic uncertainties}

The two major sources of systematic uncertainties 
affecting the results of this experiment arise from the alignment and track reconstruction 
in the FPP, and from the spin transport in the magnetic elements of the HRS. 
In addition, from Eq. \ref{eq:ratio1}, 
there is a small contribution due to the uncertainties in 
beam energy and scattered electron's angle and energy; for $Q^2$ = 3.5~GeV$^2$, the combined 
relative uncertainty is $3\times 10^{-3}$. 

In the FPP, the instrumental asymmetries cancel to first order when the difference of 
the azimuthal angular distribution is calculated for the two helicities as discussed in section III B 1 and 
shown in Figs. \ref{fig:qsqr08} and \ref{fig:qsqr35}; hence these asymmetries do not 
contribute to the systematic uncertainties.  No correction is necessary
for dead time with a polarimeter of this kind, because an event is defined in
software by the tracks in the chambers, independently of the experimental
trigger efficiency.  The dead-time is not helicity or polarization dependent as the rate 
of scattering are the same
for the two helicities, and independent of the polarization of the proton.  

However, we estimate that uncertainties of the order of 2~mrad 
in the azimuthal  angle, $\varphi$ contribute to the systematic uncertainties. This uncertainty
in $\varphi$ is due to the alignment accuracy of the FPP coordinate 
system relative to the reaction coordinate system at the target; it is based on
the accuracy with which the focal plane chambers were surveyed, and the
accuracy with which the FPP chambers were aligned relative to the focal plane 
chambers, as explained in part III A. 
  
The main contribution to the systematic uncertainties in the $G_{Ep}/G_{Mp}$ 
ratio comes from the calculation of the spin precession in the spectrometer.
The COSY model was optimized to give the same results as ESPACE when 
reconstructing the target variables $\theta $, $\delta$, $\phi$, and 
$y$ from the measured focal plane observables $\theta^{fp}$, $x^{fp}$, $\phi^{fp}$and $y^{fp}$.

The ESPACE optics matrix elements are well known from several optical 
studies of the Hall A HRSs. The uncertainties in the ESPACE reconstruction 
procedure generate uncertainties in the COSY model, and accordingly in the 
spin transport calculations. To estimate these systematics, one would have 
to vary the reconstructed target quantities and after each variation, re-optimize 
the COSY model and recalculate the spin transport matrix. The COSY 
optimization is not a unique procedure because one can achieve similar results 
by varying different model parameters. However the geometrical approximation
introduced in Section IIIB-2, given by Eq. (\ref{eq:3dipole}), shows 
that the spin transfer coefficients depend mainly on the trajectory bending 
angles Eq. (\ref{eq:chiphi}) in the dispersive and non-dispersive directions 
and much less on any other target or focal plane quantities separately, indicating 
that the details of the COSY parameterization are not important.
This approximation, in which the matrix elements are calculated from the 
trajectory bend angles, represents the trivial part of the result. 

\begin{figure}[h]
\begin{center}
\epsfig{file=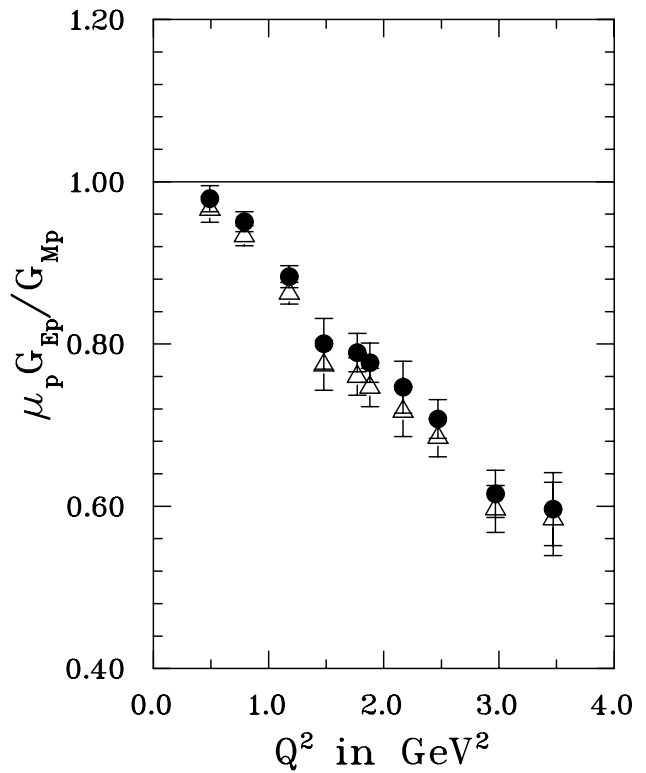,width=3.25in}
\vspace{0.15in}
\caption[]{The ratio $\mu_p G_{Ep}/G_{Mp}$ vs. $Q^2$, solid circles, obtained with 
the COSY model, compared to 
results obtained using the approximation given by Eq. (\ref{eq:3dipole}), open triangles.}
\label{fig:dipole}
\end{center}
\end{figure}

The ratios $\mu_p G_{Ep}/G_{Mp}$ obtained with the geometrical approximation given by 
Eq. (\ref{eq:3dipole}), and with the full COSY calculation, are compared in  
Fig. \ref{fig:dipole}; the difference between the results from these two methods is small and 
within the experimental uncertainty shown in Fig. \ref{fig:gepgmp_prl}.
Hence, the analysis of the experimental uncertainties can be done 
by examining the geometrical approximation given by Eq. (\ref{eq:3dipole}).
Accordingly, the error analysis is performed by evaluating uncertainties in the ratio $G_{Ep}/G_{Mp}$ 
resulting from the uncertainties in the bending angles in the 
dispersive and non-dispersive directions. Table VII gives the relative change in ratio 
at each $Q^2$ value for the given change in the bending angles.

\begin{figure}[h]
\begin{center}
\epsfig{file=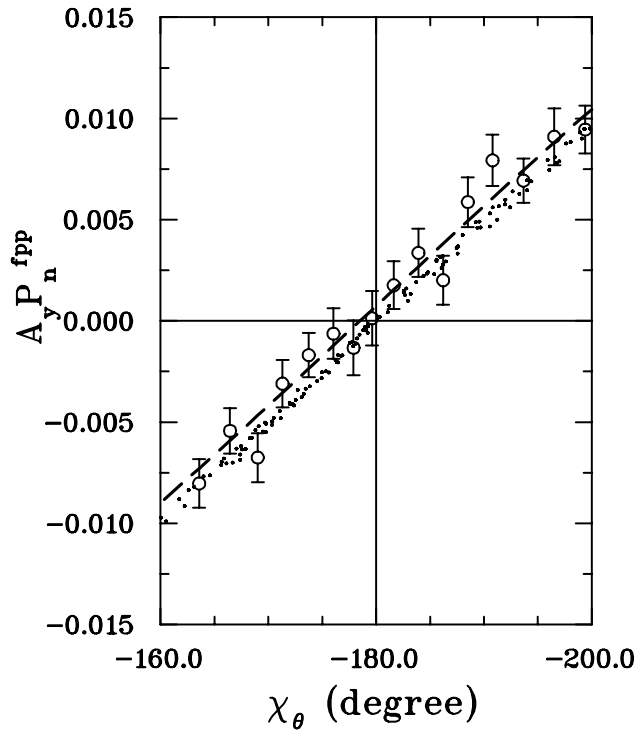,width=3.25in}
\vspace{.15in}
\caption[]{The normal polarization component at the FPP as a function of precession angle $\chi_{\theta}$.
Open circles are data, dashed line is the fit to the data and the dots are calculated with the COSY  
model. It is seen that zero crossing occur at 178.4$^{\circ}$ instead of 180.0$^{\circ}$.}
\label{fig:pn_zero}
\end{center}
\end{figure}

The value of the total bend angle in the dispersive direction 
cannot be measured directly; it was obtained from a measurement, 
using the fact that $P_{n}^{fpp}$=0 when $\sin~\chi_{\theta}$=0.
With the help of approximations given in Eq. (\ref{eq:3dipole}), $P_{n}^{fpp}$ can be 
written as:
\begin{eqnarray}
P_{n}^{fpp}=\sin~\chi_{\theta }~(-\sin ~\chi_{\phi }\cos ~\chi_{\phi}' hP_{t} + \cos ~\chi_{\phi } hP_{\ell })
\end{eqnarray}

The large acceptance of the HRS in $\chi_{\theta}$ provides a map of $P_{n}^{fpp}$ 
versus $\chi _{\theta }$. In Fig. \ref{fig:pn_zero}, the values of the normal component of 
the polarization at the FPP are shown versus the dispersive plane precession angle 
$\chi_{\theta}$ for $Q^2$ of 2.17~GeV$^2$; also shown in this figure are the 
values of the matrix elements $S_{n \ell}$ (multiplied by the average value of $P_{\ell}$) 
calculated with COSY. With a $45^{\circ}$ bend angle in the HRS, the precession angle is 
$\chi _{\theta }=180^{\circ}$ from Eq. (\ref{eq:chi}), when the spectrometer is tuned to 
the central proton momentum of 1.875~GeV/c. This procedure allows an accurate and independent 
determination of the spectrometer bend angle.  As demonstrated in this figure, the data crosses 
zero at 178.4$^{\circ}\pm$ 0.7$^{\circ}$ instead of $180.0^{\circ}$. The deviation of $1.6^{\circ}$
in precession angle, results in a 7~mrad deviation of the dispersive bend angle 
($\theta_B + \theta -\theta ^{fp}$), using Eq. (\ref{eq:chi}). A value of 7~mrad was 
included as an uncertainty in the dispersive bend angle ($\theta_B + \theta -\theta ^{fp}$). 

From a dedicated optical measurement described in Ref. \cite{tnote1},
the uncertainty in the bend angle in the non-dispersive plane, 
$\phi - \phi^{fp}$,  for the central trajectory 
was found to be 0.3~mrad. The uncertainties in the angles 
$\phi^d$ and $\Delta \phi^d $ 
given in Table VII, were calculated 
by varying the couplings 
$(\phi^d |y)$, $(\phi^d |\phi)$, 
$(\Delta \phi^d |\phi)$, and $(\Delta \phi^d | y)$, 
as discussed in Ref. \cite{tnote}. 
The individual uncertainties are given in Table VIII.   

\begin{center}
\begin{table}[hbt]
\caption{Dominant contributions to the systematic uncertainty of $\mu_p G_{Ep}/G_{Mp}$, 
which results from the variations in the quantities listed in the table. The total 
systematic uncertainties $\Delta_{sys}$ given in Table \ref{tab:results} are obtained from 
the values below in quadrature.}
\begin{ruledtabular}
\begin{tabular}{cccccc}
$Q^2$&$\theta-\theta^{fp}$&$\phi-\phi^{fp}$&$\phi^d, \Delta \phi^d$
&$\phi^{fpp}$&$\Delta_{sys}$ \\
GeV$^2$ & 7 mrad &  0.3 mrad & from Table VIII  &  2 mrad & \\ \hline
0.49 & -0.004 & 0.001 & 0.001 & -0.004 & 0.006 \\
0.79 & -0.009 & 0.001 & 0.003 & -0.004 & 0.010 \\
1.18 & -0.017 & 0.002 & 0.004 & -0.003 & 0.018 \\
1.48 & -0.025 & 0.002 & 0.004 & -0.005 & 0.026 \\
1.77 & -0.034 & 0.002 & 0.006 & -0.002 & 0.035 \\
1.88 & -0.032 & 0.002 & 0.006 & -0.005 & 0.033 \\
2.13 & -0.032 & 0.003 & 0.009 & -0.006 & 0.034 \\
2.47 & 0.030  & 0.003 & 0.010 & -0.009 & 0.033 \\
2.97 & 0.018  & 0.004 & 0.009 &  0.003 & 0.021 \\
3.47 & 0.009  & 0.005 & 0.010 & -0.001 & 0.014 \\
\end{tabular}
\end{ruledtabular}
\label{tab:syserr}
\end{table}
\end{center}

\begin{center}
\begin{table}[hbt]
\caption{Composition of the contribution in column four of Table VII.}
\begin{ruledtabular}
\begin{tabular}{cccccc}
$Q^2$ & $(\phi^d | \phi)$ & $(\phi^d | y)$ & $(\Delta \phi^d |\phi)$ & $(\Delta \phi^d | y)$& total \\
GeV$^2$ &  0.082 &  0.038 m$^{-1}$ &  0.111 &  0.105 m$^{-1}$&  \\ \hline
0.49 & 0.001 & 0.000 & -0.001 & 0.000 & 0.001  \\
0.79 & 0.003 & 0.000 & -0.001 & 0.001 & 0.003 \\
1.18 & 0.003 & 0.001 & -0.001 & 0.000 & 0.004  \\
1.48 & 0.004 & 0.001 & -0.001 & 0.001 & 0.004  \\
1.77 & 0.004 & 0.004 &  0.000 & 0.001 & 0.006  \\
1.88 & 0.004 & 0.004 &  0.001 & 0.001 & 0.006  \\
2.13 & 0.006 & 0.006 &  0.004 & 0.001 & 0.009  \\
2.47 & 0.007 & 0.003 &  0.005 & 0.000 & 0.010  \\
2.97 & 0.006 & 0.002 &  0.006 & 0.000 & 0.009 \\
3.47 & 0.004 & 0.003 &  0.007 & 0.005 & 0.010 \\
\end{tabular}
\end{ruledtabular}
\label{tab:syserr1}
\end{table}
\end{center}

\subsection{Discussion of Radiative Corrections}

No radiative corrections have been applied to the results presented in this paper.
External radiative effects are canceled by 
switching the beam helicity. The internal correction is due to hard
photon emission, two-photon exchange and higher-order contributions. 
A dedicated calculation of the radiative correction for single photon 
emission has been done by Afanasev {\it {et al.}} \cite{afanasev}. 
Their calculation includes radiative corrections to asymmetries in elastic $ep$ 
scattering for experiments in which events are selected on the basis of the 
hadronic variables only, {\it i.e.} the four-momentum transfer $Q^2$ is 
calculated from 
$(p_2~-~p_1)^2$, where $p_2$ an $p_1$ 
are the final and initial proton four-momentum respectively, and not 
from the photon 
momentum, so that the integration over this photon momentum can be 
performed analytically.  The correction to the polarization 
observables of the proton due to single photon emission is of the order 
of a few per cent and the relative correction on the polarization ratio is 
no bigger than 1{\%}. Current indications are that 
the contributions due to the other two processes mentioned above, are at the same percentage level.

\section{ Comparison to Theoretical Predictions}

The fundamental understanding of the nucleon form factors in terms of QCD 
is one of the outstanding problems in nuclear physics. So far  all 
theoretical models of the nucleon form factors 
are based on effective theories; they all rely on a comparison with existing 
data and their parameters are adjusted to fit the data. 
The much improved quality of the data from JLab has made a significant
impact on theoretical models. 

There are a number of different approaches to calculate nucleon form factors. 
Vector Meson Dominance (VMD) models~\cite{iach,hohler,gari,mergell,hammer,lomon,iachel2,bijker} 
explain low to intermediate $Q^2$ 
behavior of the form factors. Relativistic constituent quark models
\cite{chung,aznau,dziembowski,gamiller,cardarelli,pace,Desanctis,boffi,gross,walcher} 
work well in the 
$Q^2$ range of this experiment.  
Other quark models that predict nucleon form factors in this intermediate 
$Q^2$ range include the cloudy bag model \cite{lu}, di-quark model \cite{kroll,ma}, and
QCD sum rules \cite{radyushkin}. 
The soliton model of Holzwarth \cite{holzwarth} treats nucleons as extended 
objects like 
skyrmions and predicts nucleon form factors for the intermediate $Q^2$ range. 
Perturbative QCD predicts form factor values for large  
$Q^2$ \cite{brodsky,brodlep,brodsky1,belitsky}.   
In the VMD approach, the photon couples to the nucleon 
via vector mesons, whereas in QCD models the photon couples to 
the quarks directly. We discuss some of these calculations in more detail 
here and compare them with the  new data.

\subsection{ Vector Meson Dominance Models}

Early VMD model calculations of form factors included the $\rho$ and its higher excited states
for the isovector part, and the  $\omega$  and $\phi$ for the isoscalar part. The number of
mesons involved in the interaction and the coupling constants and masses of the mesons
can be varied to fit the data. 

\begin{figure}[h]
\begin{center}
\epsfig{file=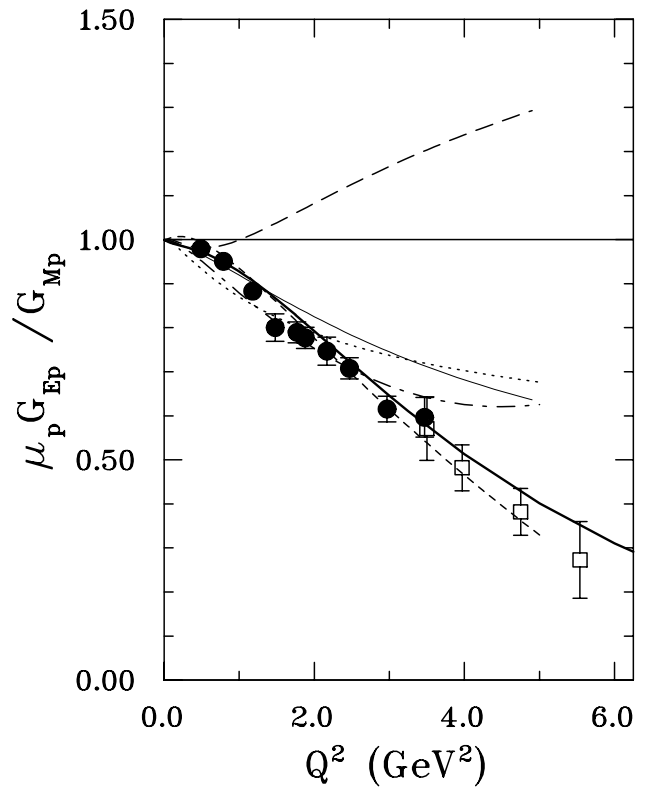,width=3.25in}
\vspace{0.1in}
\caption[]{A selection of theoretical calculations with VMD models, compared to the 
data from this experiment (solid circles) and of Ref. \cite{gayou} (empty squares). 
The curves are short dashed\cite{iach}, dot-dashed\cite{hohler}, 
thin solid and dashed\cite{gari}, dotted\cite{mergell} and thick solid\cite{lomon}.}
\label{fig:gegmvmd}
\end{center}
\end{figure}

In Fig.~\ref{fig:gegmvmd} the fits done in
the 1970s by Iachello {\it et al.} \cite{iach} and  H\"{o}hler {\it et al.} \cite{hohler} 
are shown as the short dashed line and dashed-dot line, respectively. Their predictions
come close to the data of this experiment, because the data available in 
the 1970s are compatible with the data from this experiment.

The VMD calculation of Gari and Kr\"{u}mpelmann \cite{gari} incorporated a pQCD constraint  
to extend the $Q^2$ range of the VMD description.  
Their fit to the data existing 
in 1985 is shown as a thin solid line in Fig.~\ref{fig:gegmvmd}. A new fit
was done by Gari and Kr\"{u}mpelmann  after  $G_{Ep}$ was
re-measured with smaller error bars 
in Ref.~\cite{walker} for $Q^2$ between 1.0 and 3.0 GeV$^2$; this fit is shown in Fig.~\ref{fig:gegmvmd} as
a dashed line\cite{gari}. A large difference with the earlier fit is seen; the new data from JLab clearly favor 
the older fit. 

The VMD  model of Mergell $\it {et~ al.}$
\cite{mergell} is an expansion of the original work of H\"{o}hler $\it {et~ al.}$ 
\cite{hohler}. It includes the new data from Ref. \cite{andivahis} 
and a ``super-convergence'' condition to 
constrain the behavior of the form factors to the QCD predicted fall-off at 
large $Q^2$. In this model, above $Q^2$ = 2.0 GeV$^2$, the parameter which 
indicates the boundary between mesonic and quark degrees of freedom is sensitive 
to $\mu_p G_{Ep}/G_{Mp}$ and should be tightly constrained by the new data. 
The result of their fit is shown by the dot curve in Fig.~\ref{fig:gegmvmd}. A 
recent re-examination of this model \cite{hammer} using slightly modified 
parameters succeeds at 
reproducing recent neutron form factors from polarization experiments, but still 
fails by a 
factor of 2 to reproduce the $G_{Ep}/G_{Mp}$ ratio from this work at Q$^2$=3.5 
GeV$^2$.
Lomon \cite{lomon} re-visited the Gari and Kr\"{u}mpelmann \cite{gari} approach and
fitted the form factor data including data from this 
experiment. This  model include the $\rho$, $\omega$ and $\phi$ vector meson 
pole contribution 
and pQCD constraint. Lomon also incorporated the width of the $\rho$ meson and 
higher mass vector meson 
exchanges based on the H\"{o}hler-Pietarinen \cite{hohler} model. All 
four nucleon elastic 
form factors and the ratio $\mu_p G_{Ep}/G_{Mp}$ from this experiment are fitted. The new 
fit 
for the ratio is shown in Fig.~\ref{fig:gegmvmd} as thick solid curve.
Like-wise, in 2003 Iachello and collaborators reconsidered their 1973 
work \cite{iach}; their model consists of a small structure besides the pion cloud 
described by VMD. They are now able to describe the neutron form factors better; 
and still retain the original slope for $G_{Ep}/G_{Mp}$ in the Q$^2$ range of this
experiment, although now the ratio crosses zero at Q$^2>$ 10 
GeV$^2$ \cite{iachel2}; their new work includes a prediction of the nucleon form 
factors in the time-like domain\cite{bijker}.

\subsection{Constituent Quark Models}

In the constituent quark model, the nucleon consists of three constituent quarks, which
are thought to be valence quarks dressed with gluons and quark-antiquark pairs that
are much heavier than the QCD Lagrangian quarks. All other degrees of freedom are absorbed into 
the masses of these quarks. The early success of the non-relativistic constituent 
quark model was in describing the spectrum of baryons and mesons with correct 
masses\cite{isgur}. However, to describe the data presented here in terms of constituent quarks, 
it is necessary to include relativistic effects because the momentum transfers involved
are up to 10 times larger than the constituent quark mass.     
 
\begin{figure}[h]
\begin{center}
\epsfig{file=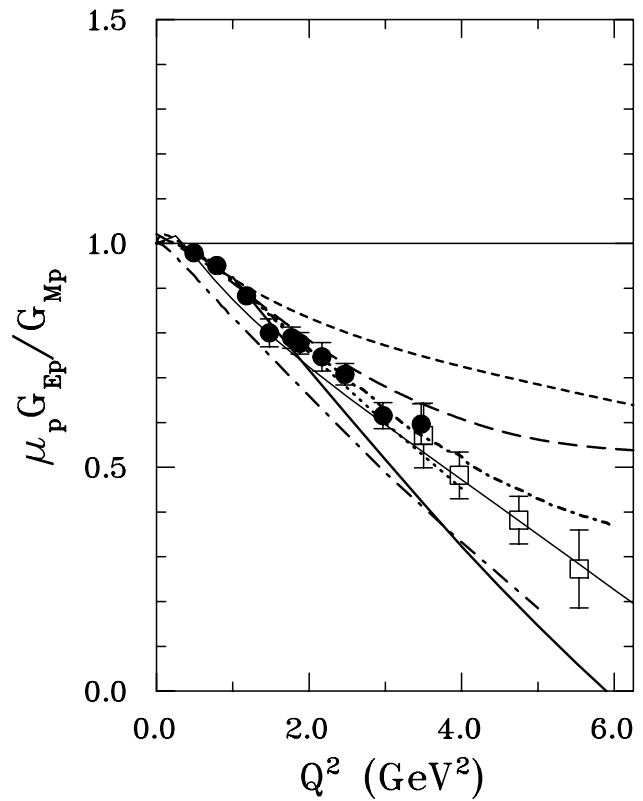,width=3.25in}
\vspace{0.1in}
\caption[]{Comparison of constituent quark model calculations with the 
data of this experiment (solid circles) and from Ref. \cite{gayou} (empty squares). 
The curves are: dotted \cite{chung}, 
thick solid \cite{gamiller}, short dot-dashed \cite{cardarelli}, 
dot-dashed and dashed \cite{pace}, short dashed \cite{boffi}, and thin solid \cite{gross}.}
\label{fig:gegmCQM}
\end{center}
\end{figure}

In the earliest study of the relativistic constituent quark models (RCQM), 
Chung and Coester \cite{chung} calculated electromagnetic nucleon form factors with 
Poincar\'{e}-covariant constituent-quark models and investigated the effect of the 
constituent quark masses, the anomalous magnetic moment of the quarks, 
and the confinement scale parameter; the prediction is shown as a dotted line in \ref{fig:gegmCQM}.

Frank $\it {et~ al.}$ \cite{gamiller} have calculated $G_{Ep}$ and $G_{Mp}$ 
in the light-front constituent quark 
model and predicted that $G_{Ep}$
might change sign near 5.6 GeV$^2$; this calculation used the light-front 
nucleonic wave function of Schlumpf \cite{schlumpf}. The light-front dynamics can 
be seen as a Lorentz transformation to a frame boosted to the speed of light. 
Under such a transformation, the spins of the constituent quarks 
undergo Melosh rotations. These rotations, by mixing spin states, play an important role 
in the calculation of the form factors. The results of their calculation are shown 
as the thick solid line curve in Fig. \ref{fig:gegmCQM}.

Several calculations with the RCQM have been motivated by the data from 
this experiment \cite{cardarelli,pace,Desanctis,boffi}. Cardarelli {\it et al.} 
\cite{cardarelli} calculated the ratio with light-front dynamics and 
investigated the effects of SU(6) symmetry breaking. They showed that the decrease 
in the ratio with increasing $Q^2$ is due to the relativistic effects generated 
by Melosh rotations of the constituent quark's spin (short dot-dashed in Fig. \ref{fig:gegmCQM}). 
In Ref. \cite{pace}, they pointed 
out that within the framework of the RCQM with the light-front formalism, an effective 
one-body electromagnetic current, with a proper choice of constituent quark form factors, 
can give a reasonable description of pion and nucleon form factors. The results of their calculation with two 
different quark form factors are shown as the dot-dashed and dashed curves in Fig. 
\ref{fig:gegmCQM}.
De Sanctis {\it et al.} \cite{Desanctis} have calculated the ratio $G_{Ep}/G_{Mp}$ within the 
hypercentral 
constituent quark model including relativistic corrections: however, the slope of their 
$G_{Ep}/G_{Mp}$ ratio is too small by a factor of $\sim$2.  The chiral constituent quark model 
based on Goldstone-boson-exchange dynamics was used by Boffi {\it et al.} \cite{boffi}
to describe the elastic electromagnetic and weak form factors. They compute these form 
factors in a 
covariant framework using the point-form approach to relativistic quantum mechanics. The 
results of these 
calculations are shown as the short dashed curve in Fig. \ref{fig:gegmCQM}. 
      
More recently Gross and Agbakpe \cite{gross} revisited the RCQM imposing that the
constituent quarks 
become point particles as Q$^2\rightarrow\infty$ as required by QCD; using a covariant spectator model which
allows exact handling of all Poincar\'{e} transformations, and 
monopole form factors for the constituent quarks, they obtain excellent ten parameter fits to all four 
nucleon form factors (shown as thin solid line in Fig. \ref{fig:gegmCQM}); they conclude that the 
recoil polarization data can be fitted with a 
spherically symmetric state of 3 constituent quarks. 

For inclusiveness we mention an interpretation of the nucleon form factors in terms of a quark cloud and 
constituent quarks proposed by Friedrich and Walcher~\cite{walcher}. Following on their 
observation that all four form factors
show an enhancement near Q$^2$=0.3 GeV$^2$. Their parameterization results in a very good fit to the 
$G_{Ep}/G_{Mp}$ up to the maximum Q$^2$=5.6 GeV$^2$ of the data of Ref.~\cite{gayou}.

\subsection{More Theoretical Models}

\begin{figure}[h]
\begin{center}
\epsfig{file=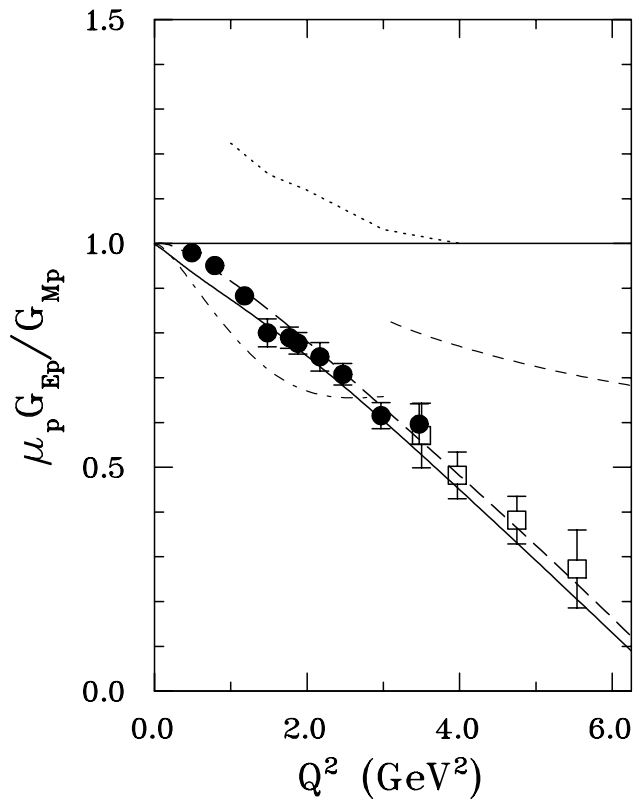,width=3.25in}
\vspace{0.1in}
\caption[]{A selection of theoretical calculations compared to the 
data from this experiment (solid circles) and from Ref. \cite{gayou} (empty squares). 
The curves are, dot-dashed \cite{lu}, short-dashed \cite{krollpc}, and dashed and solid \cite{holzwarth}.}
\label{fig:gegmotm}
\end{center}
\end{figure}

The version of the cloudy bag model (CBM) used by Lu et al. \cite{lu} couples a pion field to the quarks 
inside a bag such that chiral symmetry is restored. This model is an improvement over the MIT bag model. 
In CBM, quarks are confined within a finite spherical well of radius R of about 0.8 to 1 fm. The model
wave function for a proton is the combined wave function of the individual quarks.  
As shown in Fig.~\ref{fig:gegmotm} by the dashed-dot curve, the prediction is low compared to the
data from this experiment, but this is a simplified version of the model.
The authors suggest that one possible future development would be
to include $\pi\pi$ interactions.

Predictions of the nucleon form factors were done  by Kroll {\it et al.} \cite{kroll} in a 
framework where
the nucleon consists of quark and diquark constituents. The diquark has a finite size and
its composite nature is taken into account by form factors.  This model describes the proton
in terms of distribution amplitudes, and the photons and gluons couple to the diquark as 
in perturbative QCD; hence in the limit of
\( Q^{2}\rightarrow \infty  \) the diquark picture becomes the hard
scattering formalism of pQCD. The parameters of the model
were determined by fits to the $ep$ cross sections above $Q^2$ = 3.64 GeV$^2$.
Two sets of parameters gave equally good fits to the $ep$ cross sections, but 
differed markedly in their predictions for $G_{Ep}$ and the neutron form
factors. The diquark-quark model is not expected to be applicable below $Q^2$ 
= 3 GeV$^2$. A recent prediction \cite{krollpc} from this model is shown in 
Fig.~\ref{fig:gegmotm} as the medium dashed curve. Another study of the di-quark model with
connection to the generalized parton distribution is discussed in Ref.~\cite{tiburzi}. It
describes the proton in the double distribution formalism, as a bound state state of a residual quark
and two quarks strongly coupled in the scalar and axial-vector diquark channel. The work is currently 
limited to the small Q$^2$ region ($<$ 0.3 GeV$^2$).       

In Fig.~\ref{fig:gegmotm} we also show the QCD sum rule based prediction of 
Radyushkin \cite{radyushkin} (dot curve). This approach is based on the quark-hadron
duality concept, which assumes that the characteristics of the free or almost free quarks in 
perturbation theory are close to the analogous characteristics of the hadronic spectrum 
integrated over an appropriate energy region. The validity of this model is in the very large
$Q^2$ region.

In the soliton model Holzwarth \cite{holzwarth} applies the relativistic corrections 
due to recoil and incorporates partial coupling to vector mesons. He
uses the skyrmion as an extended object with one vector meson propagator and relativistic 
boost to the Breit frame. The result is shown in Fig.~\ref{fig:gegmotm} as the dashed and solid curves, 
corresponding to two different strength of the $\omega$-meson coupling strength, $g_{\omega}$. This
model describes the ratio very well over the Q$^2$ range of this experiment.  
 
\subsection{pQCD Predictions}

\begin{figure}[h]
\begin{center} 
\epsfig{file=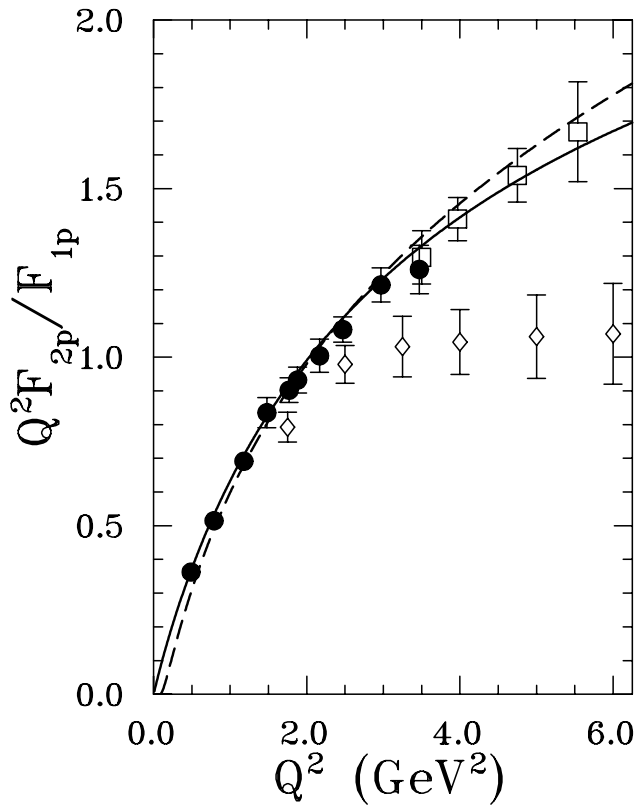,width=3.25in}
\vspace{0.1in}
\caption[]{$Q^2 F_{2p}/F_{1p}$ versus $Q^2$. The curves shown are from Brodsky \cite{brodsky1} (solid line), 
and from Belitsky {\it et al.} \cite{belitsky} (dashed line). The data from this experiment are shown as 
solid circles, from Ref. \cite{gayou} as  empty squares and from Ref. \cite{andivahis} as empty diamonds.}
\label{fig:q2f2f1}
\end{center}
\end{figure}

In the pQCD approach proposed by Brodsky and collaborators\cite{brodsky,brodlep}, 
the interaction is described as a convolution 
of a hard scattering amplitude and a baryon distribution amplitude. Up to
leading order in 1/$Q^2$, the magnetic form factor is proportional to 
$\alpha_s/Q^4$ times slowly varying logarithmic terms, because the momentum of the virtual
photon absorbed by one quark, must be shared with the two other quarks through the exchange of
two gluons. In lowest order the absorbed virtual photon cannot induce a quark helicity flip and 
consequently pQCD predicts a faster decrease of $F_{2p}$ 
than of $F_{1p}$ with increasing $Q^2$, by a factor $Q^{-2}$ \cite{brodsky}. 
The expectation is then that $Q^2 F_{2p}/F_{1p}$ should
become constant at very high $Q^2$. 
In Fig.~\ref{fig:q2f2f1} the JLab data together with data of Andivahis {\it et al.}\cite{andivahis},
are shown as $Q^2 F_{2p}/F_{1p}$. The data from Ref. \cite{andivahis} show flattening 
above $Q^2$ of 3 GeV$^2$. However, the data from this experiment do not show yet the specific pQCD $Q^2$ dependence.

Recently there have been  two revisions of the pQCD prediction
for the large $Q^2$ behavior of $F_2$. In the first, Brodsky \cite{brodsky1} argues 
that the pQCD motivated behavior of $F_2$ must contain an extra logarithmic 
term from higher twist contributions; the 3 free parameters $a$, $b$ and $c$ of the expression
\begin{equation}
\frac{F_{2p}}{F_{1p}}=\frac{1}{1+(Q^{2}/c)\ln^b(1+Q^{2}/a)}
\end{equation}
\noindent
were fitted in Ref. \cite{brodsky1} to the data 
presented here augmented from the data of Ref. \cite{gayou} with the result shown 
as a solid line in Fig. \ref{fig:q2f2f1}. In the second, Belitsky $\it {et~ al.}$ \cite{belitsky} 
reiterate the fact that the spin of a massless (or very light) quark 
cannot be flipped by the virtual photon of the $ep$ reaction. For a quark to 
undergo spin-flip, it must be in a state of non-zero angular momentum with projection $\mid L_z \mid = $1. 
As a result, the standard pQCD prediction for $F_{2p}$ (namely $\propto Q^{-6})$ 
becomes modified by a logarithmic term such that:
\begin{equation} 
\frac{F_{2p}}{F_{1p}}=\frac{A } {\kappa_{p} Q^2} \ln^2(\frac{Q^2}{\Lambda^2}),
\end{equation} 
\noindent
where $A$ is a 
normalization constant;    
$\Lambda$ is a cutoff constant required to suppress the infrared singularity
generated by the very soft part of the quark wave function. Although the
constant $A$ in the expression above is not determined, a fit to the data 
of this paper augmented by the data of \cite{gayou} gives $\Lambda=290$~MeV, and
$A=$0.175. The soft physics scale of the nucleon is determined by $\Lambda$; its 
size is of order of the transverse quark momentum in the nucleon. 
This fit is shown as the dashed line in Fig. \ref{fig:q2f2f1}. 

Polynomial fits to all four Sachs form factors by Kelly \cite{kelly,kelly1} indicate
that the data from this experiment are compatible with an approach to the pQCD regime 
following dimensional scaling.
 
\begin{figure}[h]
\begin{center}
\epsfig{file=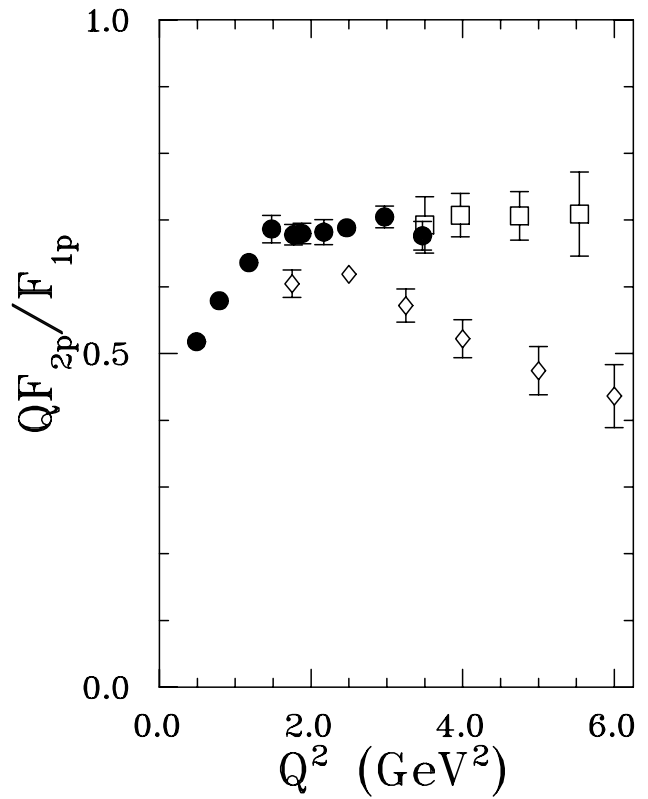,width=3.25in}
\vspace{0.1in}
\caption[]{The data from this experiment, from Ref. \cite{gayou}, and \cite{andivahis} are shown as 
$Q F_{2p}/F_{1p}$ vs.$Q^2$.}
\label{fig:qf2f1}
\end{center}
\end{figure}

In Fig.~\ref{fig:qf2f1} the JLab data plotted as
$QF_{2p}/F_{1p}$ show a remarkable flattening of the ratio starting at 1-2 GeV$^2$. 
Inspired by the results of this experiment,  Ralston \cite{ralston,jain} revisited 
the calculation of the single-quark spin-flip 
amplitude responsible for the Pauli form factor in the framework of QCD and
conclude that if quarks in the proton carry orbital angular momentum, 
then  $F_{2p}/F_{1p}$ should behave like $\frac{1}{\sqrt{(Q^2)}}$, 
rather than the well known pQCD prediction of ${1}\over{Q^2}$ (Ref. \cite{brodsky}). 
In a different approach, Miller and Frank ~\cite{miller} 
have shown that imposing Poincar\'{e} invariance leads to violation of the helicity 
conservation rule, which results in the behavior of $F_{2p}/F_{1p}$ observed in the JLab data.

\subsection{Generalized Parton Distributions, Lattice QCD and Form Factors in the Time Like Region}

The generalized parton distributions (GPDs) provide a framework to describe the process of emission 
and re-absorption of a quark in the non-perturbative region by a hadron in exclusive reactions;
they are universal non-perturbative objects describing hard exclusive processes induced by photon 
and electrons or positrons. 
Precise measurements of elastic nucleon form factors provide stringent constraints 
on the parameterization of the GPDs. Early theoretical developments in GPDs indicated that 
measurements of the separated elastic form factors of the nucleon to high $Q^2$ may shed
light on the problem of nucleon spin. The first moment of the GPDs taken
in the forward limit yields, according to the Angular Momentum Sum
Rule \cite{ji}, a contribution to the nucleon spin from the quarks and gluons, 
including both the quark spin and orbital angular momentum.  The $t$-dependence of the GPDs
has been modeled using a factor corresponding to the relativistic Gaussian 
dependence of both Dirac \cite{rady} and Pauli \cite{afa1} form factors of 
the proton. Extrapolation of these GPDs to $t$ = 0 leads to the functions entering
into the Angular Momentum Sum Rule, and an estimate of the contribution
of the valence quarks to the proton spin can then be obtained. This approach 
was used by Afanasev \cite{afan} who, using the $G_{Ep}/G_{Mp}$ data of this experiment
in the framework of GPDs, concluded that valence quarks contribute about 50 \% of the nucleon 
spin. When combined with inclusive deep inelastic scattering data from SMC \cite{ashman}, 
this result implies that about 25 \% of the proton spin comes from the orbital angular momentum 
of the valence quarks. In a recent development of these ideas Guidal {\it et al}
~\cite{guidal} have shown that 
the difficulties encountered in the Gaussian parameterization used in earlier work could be 
surmounted with a Regge parameterization. In one version of their parameterization, these authors
obtain excellent fits to all 4 nucleon form factors. In a different approach Diehl 
{\it et al}~\cite{diehl}
use theoretically motivated parameterizations of the relevant GPDs in the very small and very large 
x-domains, and interpolated by fitting the nucleon Dirac form factors $F_1^p$ and $F_1^n$. They derive the valence contribution to Ji's sum rule~
\cite{ji}. In a related approach, Ji~\cite{ji2} has shown that the GPDs provide a classical 
visualization of the quark orbital motion. 

There are also lattice QCD calculations predicting the
contribution to the proton spin coming from angular momentum of the
valence quarks. For example, Mathur {\it {et al.}} \cite{mathur} have calculated the
quark orbital angular momentum of the proton from the quark energy-momentum tensor form factors on 
the lattice,  and found the total contribution
to the proton spin from the quarks to be 60 \%,
of which 35 \% originates from the orbital angular momentum and 25 \% from the spin of the quarks.  
Most recent calculations of the GPDs from quenched lattice QCD give $u$ and $d$ valence quark
contributions to Ji's sum rule comparable to the results in Ref.~\cite{guidal}. A quenched lattice
QCD calculation by Ashley {\it et al}~\cite{ashley} with extrapolation to the quark mass gives
good agreement with $G_{Ep}$ up to Q$^2$=1 GeV$^2$; it may be a harbinger of the quality of future 
lattice calculations.  
In contrast with all the models discussed above, lattice QCD will ultimately predict form factors 
from first principles.
 
In a more general framework, the asymptotic behavior of both space-like and 
time-like elastic form factors are connected. Application of the 
Phragm\'{e}n-Lindelh\"{o}f theorem~\cite{bilenky} to the form factors shows 
that their asymptotic behavior must be the same, and that 
their ratio should go to 1 as $Q^2 \rightarrow \infty$.
The existing time-like proton form-factor data show that this condition is 
far from being satisfied at $Q^2$=13 GeV$^2$ as discussed by 
Tomasi-Gustafsson and Rekalo \cite{tomasi}. However, no form factor separation 
has been possible to this point, and therefore comparison of asymptotic 
behavior in the time- and space like domains are currently limited by the 
data base.

\subsection{Charge and Magnetization Distribution}

\begin{figure}[h]
\begin{center} 
\epsfig{file=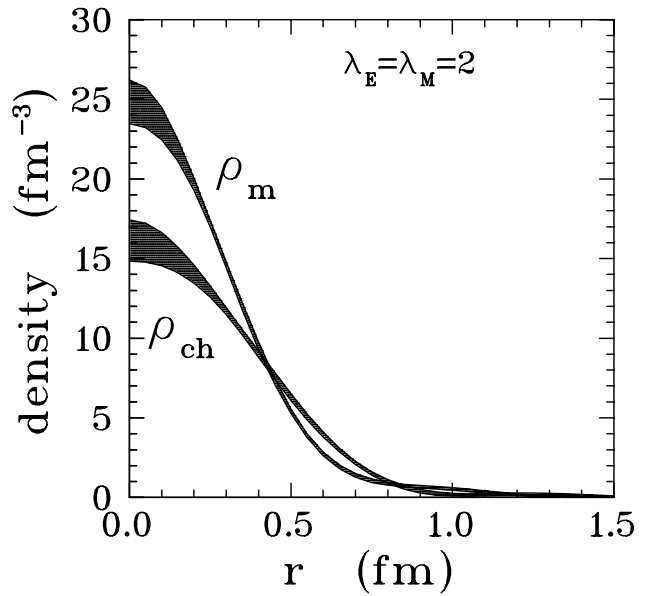,width=3.25in}
\vspace{0.1in}
\caption[]{The charge and magnetization distribution obtained by \cite{kelly} from the data of this 
experiment and of Ref. \cite{gayou}.} 
\label{fig:density}
\end{center}
\end{figure}

Recently Kelly \cite{kelly} has developed a relativistic prescription to relate  the Sachs form factors 
to the nucleon charge and magnetization densities which is consistent with pQCD at large $Q^2$ and with 
the Lorentz contraction of the Breit frame relative to the rest frame. The model dependence of 
the fitted densities is minimized by using an expansion in a complete set of radial basis functions.
Details of the fitting procedure and data selection are in Ref. \cite{kelly}; 
for $Q^2 >$ 1~GeV$^2$ the $G_{Ep}$ analysis relied upon the recoil polarization data 
from Jefferson Lab and omitted Rosenbluth separation data. Fig. \ref{fig:density} compares 
the fitted charge and magnetization densities, where both densities are normalized such that 
$\int_{0}^{\infty} \rho(r)~ r^2~dr~=~1$ and where the bands include both statistical and 
incompleteness errors. The charge density is significantly broader than the magnetization because $G_{Ep}$ is
softer than $G_{Mp}$, falling more rapidly with respect to $Q^2$.  

\section{CONCLUSIONS}

We have presented the results of a measurement of the proton elastic form 
factor ratio, $G_{Ep}/G_{Mp}$, obtained in a polarization transfer experiment
at JLab.  The ratio values decrease systematically with increasing 
four-momentum  transfer squared, $Q^2$, starting at about 1~GeV$^2$ and up to 
3.5~GeV$^2$. Comparing the new data with the database of all previous experiments,
we conclude that the interpretation of the previous database, widely 
accepted in the literature, that 
the $G_{Ep}/G_{Mp}$ ratio is approximately constant up to 5~GeV$^2$, is no longer 
sustainable.   
As the discussion in this paper explains, the great strength of the 
polarization transfer method is the tight control of systematics it 
affords. Previous arguments about the difficulty to separate $G_{Ep}$ from 
cross section data were based on the expectation that the ratio 
was essentially $1/\tau\mu_p$; but the polarization results indicate that 
already at 3.5~GeV$^2$ the ratio has decreased to 60 \% of this 
expectation. The consequence 
is of course that at $Q^2$ larger than 3.5~GeV$^2$, the power of the polarization
method will become even more obvious. A continuation of this experiment to
larger $Q^2$, using the same technique and instrumentation, has recently been 
published \cite{gayou}; it shows that the ratio continues to decrease, 
reaching a value of 0.27 at $Q^2$=5.6~GeV$^2$ (see Fig. \ref{fig:gepgmp_prl}). 

There are new ratio results from two JLab experiments \cite{christy,segel} obtained 
with the Rosenbluth separation method; these results agree with previous Rosenbluth results 
\cite {litt,berger,price,walker,andivahis} and confirm 
the discrepancy between the ratios obtained with the Rosenbluth separation method and 
the recoil polarization method. The origin of this discrepancy is currently the subject of
intense discussion. A likely explanation is the two-photon exchange process, which affects 
both cross section and polarization transfer components at the level of a few percents. 
However, the two-photon effects have drastic effect on the results from a Rosenbluth 
separation method, but modifies the ratio obtained with polarization method by a few percent
only. A new experiment at JLab \cite{lubomir} will investigate the two-photon effects 
in the near future.      

Theoretical calculations of the proton electromagnetic form factors have a 
long history. The database for three out of the four nucleon form factors is limited 
to 10~GeV$^2$ or less, reaching 31.2~GeV$^2$ only for $G_{Mp}$. The basic physics of the 
interaction of the electromagnetic probe with the nucleon is in the difficult
region of transition between pictures of the nucleon, as a small core surrounded 
by a meson cloud, and a system of three valence quarks accompanied by gluons 
and quark-antiquark pairs described by QCD. At the lower end of this $Q^2$ range, 
the assumption that the photon interacts predominantly via an intermediate 
vector meson has been very successful; recent reconsideration of this model provides 
a quantitative description of 
the data for all four form factors. Below $Q^2$ of 10~GeV$^2$, one must use non-perturbative QCD, 
and only QCD-based phenomenological models of the nucleon are 
available. The most successful QCD based model is 
the relativistic constituent quark model, 
which describes the drop-off in the ratio $G_{Ep}/G_{Mp}$ observed 
in this experiment.    
At a very large, but not quantitatively defined $Q^2$, a perturbative version of 
QCD (pQCD) pioneered by Brodsky and collaborators \cite{brodsky,brodlep} should 
be valid. An important 
consequence of pQCD is hadron helicity conservation; in terms of the 
non-spin flip and spin flip  form factors (Dirac and Pauli), pQCD has
generally been understood to predict a faster decrease with $Q^2$ for $F_{2p}$ 
than $F_{1p}$, by a factor of 1/$Q^2$. The data presented here clearly show that
the ratio $Q^2F_{2p}/F_{1p}$ is still increasing monotonically up to 3.5~GeV$^2$. 
Recently a careful re-examination of the pQCD prediction has led to the inclusion of 
a logarithmic factor and good agreement with the behavior of $F_{2p}/F_{1p}$ reported in this paper.

\section{ACKNOWLEDGMENTS}

The collaboration thanks the Hall A technical staff and the Jefferson Lab 
Accelerator Division for their outstanding support during this experiment. 
This work was supported by DOE contract DE-AC05-84ER40150 Modification No. M175, 
under which the Southeastern Universities Research Association (SURA) operates 
the Thomas Jefferson National Accelerator Facility. 
This work was supported in part by the U.S. Department of Energy, the U.S. 
National Science Foundation, the 
Italian Istituto Nazionale di Fisica Nucleare (INFN), the French Commissariat 
\`a l'Energie Atomique, Centre National de la Recherche Scientifique 
(CNRS), and Conseil R\'egional d'Auvergne, and the Natural Sciences and Engineering 
Research Council of Canada. The polarimeter was funded by U.S. National Science Foundation
grants PHY 9213864 and 9213869.

\section{Appendix A}

\subsubsection{Old Method}
  
Using Eq. (\ref{eq:azimuth})
we construct the asymmetry $f(\varphi)$ and the efficiency function $E(\varphi)$ as:

\begin{eqnarray}
f (\varphi)& = & (f^+(\varphi )-f^-(\varphi ))/2  \nonumber \\
&=& \frac{\epsilon (\varphi )}{2 \pi } \Big[A_yP_t^{fpp}\sin \varphi 
-A_yP_n^{fpp}\cos \varphi \Big] 
\label{eq:assymf}\\
E(\varphi)& = & (f^+(\varphi )+f^-(\varphi ))/2 =\frac{\epsilon(\varphi)}{2 \pi} 
\label{eq:assymf1}
\end{eqnarray}
\noindent
In the next step, using Eq. (\ref{eq:pnpt}) we replace $P_t^{fpp}$ and $P_n^{fpp}$ to obtain following relations:

\begin{eqnarray}
f(\varphi)&=& \frac{\epsilon(\varphi)}{2\pi} \Big[hA_yP_tS_{tt}\sin\varphi+hA_y P_{\ell} S_{t \ell} \sin\varphi \nonumber\\ 
&-& hA_yP_t S_{nt} \cos \varphi - hA_y P_{\ell} S_{n \ell} \cos \varphi \Big]  \nonumber\\ 
&=&\frac{\epsilon (\varphi )}{2 \pi } \Big[hA_yP_t\lambda _t(\varphi)- hA_yP_{\ell} \lambda _{\ell}(\varphi) \Big]
\label{eq:assymfp}
\end{eqnarray}

\noindent

where, 

\begin{eqnarray}
\lambda _t(\varphi)&=&-S_{nt}\cos \varphi + S_{tt}\sin \varphi \nonumber \\
\lambda _\ell(\varphi)&=&S_{n\ell }\cos \varphi - S_{t\ell }\sin \varphi .
\label{eq:lambda}
\end{eqnarray}

\noindent
The Fourier analyses of the asymmetry $f(\varphi)$ leads to the following expressions:

\begin{eqnarray}
\int_0^{2\pi }f(\varphi )\cos \varphi  d\varphi&=& \nonumber \\ 
&=&hA_yP_t \int_0^{2\pi} E(\varphi) \lambda _t(\varphi) \cos \varphi d\varphi  \nonumber\\ 
&-&hA_yP_{\ell} \int_0^{2\pi} E(\varphi) \lambda _\ell(\varphi) \cos \varphi d\varphi, \nonumber\\
& & 
\label{eq:csoper}
\end{eqnarray}
\noindent
and
\begin{eqnarray} 
\int_0^{2\pi }f(\varphi )\sin \varphi d\varphi &=& \nonumber \\
&=&hA_yP_t \int_0^{2\pi} E(\varphi) \lambda _t(\varphi) \sin \varphi d\varphi  \nonumber\\
&-& hA_yP_{\ell }\int_0^{2\pi} E(\varphi) \lambda _\ell(\varphi) \sin \varphi d\varphi . \nonumber\\
& &
\label{eq:csoper2}
\end{eqnarray}
\noindent

To obtain polarization components at the target, the integrals above are approximated by 
corresponding sums over the observed events:
\begin{eqnarray}
\sum_{ev} \pm \cos \varphi &=& hA_yP_t \sum_{ev} \lambda _t(\varphi) \cos \varphi  \nonumber\\ 
&-& hA_yP_{\ell} \sum_{ev} \lambda _\ell(\varphi) \cos \varphi 
\label{eq:meth1}
\end{eqnarray}
\begin{eqnarray}
\sum_{ev} \pm \sin \varphi &= & hA_yP_t \sum_{ev} \lambda _t(\varphi) \sin \varphi  \nonumber\\
&-&  A_yP_{\ell} \sum_{ev} \lambda _\ell(\varphi) \sin \varphi ,
\label{eq:methI}
\end{eqnarray}
\noindent

In the above event sums, the 
spin matrix elements $S_{ij}$ that enter in $\lambda _{\ell ,t}(\varphi)$, 
and $\varphi$ are different for each event; they are calculated event by event and they are functions
of the target quantities $x$, $\theta $, $y $, $\phi $ and $\delta $.

\subsubsection{New Method}

In the new method we multiply the asymmetry $f(\varphi)$ in Eq. (\ref{eq:assymfp}) 
by $\lambda _{\ell} ( \varphi )$ and $\lambda _t ( \varphi )$, instead of multiplying it by 
$\cos \varphi $ and  $\sin  \varphi $ as in the old method, and integrate it over $\varphi $ as shown below:  

\begin{eqnarray}
\int_0^{2\pi}f(\varphi )\lambda _{\ell} ( \varphi ) d\varphi &=& \nonumber\\
&=&hA_yP_t \int_0^{2\pi} E(\varphi)\lambda _t ( \varphi ) \lambda _{\ell} ( \varphi ) d\varphi \nonumber\\
&-&hA_yP_{\ell} \int_0^{2\pi} E(\varphi) \lambda _{\ell}^2 ( \varphi ) d\varphi , \nonumber\\
& &
\label{eq:csoper3}
\end{eqnarray}
and 
\begin{eqnarray}
\int_0^{2\pi}f(\varphi) \lambda _t ( \varphi ) d\varphi \nonumber\\
&=&hA_yP_t\int_0^{2\pi} E(\varphi) \lambda _t^2( \varphi )  d\varphi \nonumber\\
&-&hA_yP_{\ell}\int_0^{2\pi} E(\varphi) \lambda _{\ell} ( \varphi ) \lambda _t ( \varphi ) d\varphi. \nonumber\\
& &
\label{eq:csoper1}
\end{eqnarray}

As in the old method, the polarization components at the target can then be obtained by 
replacing the integrals in Eqs. (\ref{eq:csoper3},\ref{eq:csoper1}) with corresponding sums over the observed events:

\begin{eqnarray}
\sum_{ev} \pm \lambda _{\ell} ( \varphi ) &=& \nonumber\\ 
&=&hA_yP_t\sum_{ev} \lambda _t ( \varphi ) \lambda _{\ell} ( \varphi ) \nonumber\\ 
&-&hA_yP_{\ell}\sum_{ev} \lambda _\ell^2 ( \varphi ), 
\label{eq:methII1}
\end{eqnarray}
and
\begin{eqnarray}
\sum_{ev} \pm  \lambda _t ( \varphi ) &=& \nonumber\\
&=&hA_yP_t\sum_{ev} \lambda _t^2( \varphi )  \nonumber\\
&-&hA_yP_{\ell}\sum_{ev} \lambda _{\ell} ( \varphi ) \lambda _t ( \varphi ).
\label{eq:methII}
\end{eqnarray}

This matrix equation gives the target quantities $hA_yP_{\ell}$ and $hA_yP_t$.

\end{document}